# Describing the movement of molecules in reduced-dimension models.

Natasha S. Savage.

University of Liverpool, Liverpool, UK.

**Email:** nsavage@liverpool.ac.uk

## Abstract

When addressing spatial biological questions using mathematical models, symmetries within the system are often exploited to simplify the problem by reducing its physical dimension. In a reduced-dimension model molecular movement is restricted to the reduced dimension, changing the nature of molecular movement. This change in molecular movement can lead to quantitatively and even qualitatively different results in the full and reduced systems. Within this manuscript we discuss the condition under which restricted molecular movement in reduced-dimension models accurately approximates molecular movement in the full system. For those systems which do not satisfy the condition, we present a general method for approximating unrestricted molecular movement in reduced-dimension models. We will derive a mathematically robust, finite difference method for solving the 2D diffusion equation within a 1D reduced-dimension model. The methods described here can be used to improve the accuracy of many reduced-dimension models while retaining benefits of system simplification.

## Introduction

Simple models of complex systems are an invaluable tool for gaining conceptual insight into biological mechanism [1, 2]. In spatial models a powerful simplification technique often used is to exploit symmetries within a biological system's geometry and patterning to reduce the physical dimension of the problem. For example consider a single cell and the formation of a concentrated patch of membrane associated proteins, a polarity patch, this system has radial symmetry and so mechanisms controlling patch formation could be explored by considering a one dimensional (1D) slice through the centre of the patch, rather than considering the entire two dimensional (2D) membrane [3-8], Fig. 1A. When addressing questions which include cytoplasmic gradients, for example, a system's dimension is often reduced from three to 2D [1, 9-13], or even 1D [1, 2, 14-16]. An analogy for reduced-dimension models is the focal plane in microscopy: Analysis is performed on data acquired from a representative slice through the system, then results are inferred onto the entire cell or tissue. All spatial models contain a description of molecular movement. Molecular movement in reduced-dimension models is restricted to the focal plane, compare Fig. 1B-C. Thus, when one reduces the dimension of a system and restricts molecular movement to the reduced dimension, they are changing the geometry of the problem. It is understood that cell geometry influences molecular movement and patterning [11, 13, 19-21] , SI Appendix 10. Here we present a general methodology that enables a reduced-dimension model to take the system's full geometry into account, by using it to estimate the molecular flux through the focal plane. We go on to use this general methodology to derive a simple numerical method (the 1D-uFDM) which can be used to solve the 2D diffusion equation in a 1D reduced-dimension model.

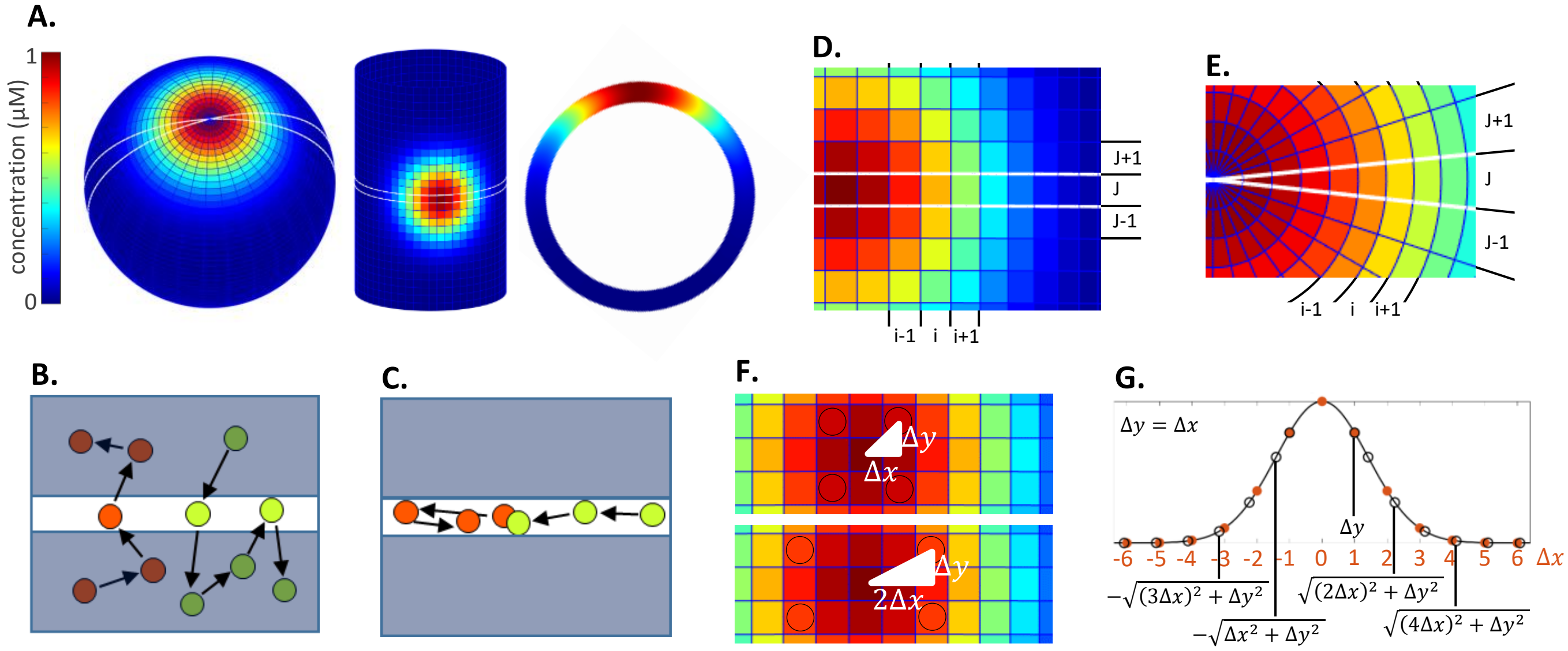

**Figure 1. (A)** Dimension reduction example showing a polarity patch on a spherical cell and body of an elongated cell with their reduced-dimension model representation, a stripe on a ring. White lines on 2D surfaces show the focal plane. **(B)** Molecules on a 2D surface moving through the focal plane. The focal plane is shown as a white stripe. **(C)** Molecules in a reduced dimension model are restricted to the focal plane (white stripe). **(D)** Regular mesh over a polarity patch on the body of an elongated cell, the zero-flux assumption is not satisfied. **(E)** Spherical mesh over a polarity patch on a spherical cell. The zero-flux assumption is satisfied, $u^{\tau}_{i,J-1} = u^{\tau}_{i,J} = u^{\tau}_{i,J+1}$. **(F)** Regular mesh over a polarity patch on the body of an elongated cell. Distances from the centre of the patch of each point in rows $J \pm 1$ found using Pythagoras (white triangles and text). Concentrations in circles are the same because of radial symmetry. **(G)** Using interpolation (black line) on the concentrations in the focal plane (orange dots) to estimate $\tilde{u}^{\tau}_{i,J\pm1}$ (black open circles).

## Results

Without loss of generality, within this manuscript, we will consider the full system to be a 2D surface, a membrane, and the reduced-dimension model to be a 1D ring (Fig.1A). Molecular movement will be diffusive. Throughout the manuscript 2D and 1D solutions are compared to analyse and illustrate the accuracy of the reduced-dimension method being proposed. A user of this method would not generate the full dimension model, they would only generate the simplified reduced-dimension model.

### Reduced-dimension models have a zero-flux assumption

To investigate any inbuilt assumptions of reduced-dimension models we compared the finite difference solution to the diffusion equation in 2D and 1D. For the 2D solution cover the 2D membrane with a regular mesh, Fig. 1D. The mesh-points along the $x$-axis are labelled $i = 1,2, \dots, N$ and are distance $\Delta x\ \mu m$ apart, the mesh-points along the $y$-axis are labelled $j = 1,2, \dots, M$ and are distance $\Delta y\ \mu m$ apart. Let $u^{\tau}_{i,j}$ represent the concentration of molecule $u$ on the membrane at point $(i, j)$ and time $\tau$. The change in concentration of $u^{\tau}_{i,j}$ over time, as a result of diffusive movement, is described by the solution to the 2D diffusion equation. The explicit 2D finite difference scheme (2D-FDM) used to solve the 2D diffusion equation on a regular mesh, with diffusion coefficient $D\ \mu m^2 s^{-1}$, is [17],

$$u_{i,j}^{\tau+1} = u_{i,j}^{\tau} + \frac{\Delta t}{\Delta x^2} D\left(u_{i-1,j}^{\tau} - 2u_{i,j}^{\tau} + u_{i+1,j}^{\tau}\right) + \frac{\Delta t}{\Delta y^2} D\left(u_{i,j-1}^{\tau} - 2u_{i,j}^{\tau} + u_{i,j+1}^{\tau}\right) \qquad \text{explicit 2D-FDM}$$

Note that to find $u_{i,j}^{\tau+1}$ the 2D-FDM calculates molecular movement along the $x$-axis and $y$-axis separately.

To build a 1D reduced-dimension model describing the diffusive movement of $u$ on a 2D surface, the focal plane is set to run through the axis of symmetry of $u$ and the membrane, Fig. 1A. Molecule movement through the focal plane is then approximated by the solution to the 1D diffusion equation. Recall the 2D mesh, assume that the focal plane is set along the $x$-axis at row $j = J$, Fig. 1D. The explicit 1D finite difference scheme (1D-FDM) used to solve the 1D diffusion equation is [17],

$$u_{i,J}^{\tau+1} = u_{i,J}^{\tau} + \frac{\Delta t}{\Delta x^2} D\left(u_{i-1,J}^{\tau} - 2u_{i,J}^{\tau} + u_{i+1,J}^{\tau}\right) \qquad \text{explicit 1D-FDM}$$

As the 1D reduced-dimension model only contains information about concentrations on row $J$ the 1D-FDM contains no terms for calculating the movement of $u$ along the $y$-axis (compare the 2D and 1D-FDMs). Thus, molecular movement in the 1D reduced-dimension model is modelled as though it is restricted to the focal plane Fig. 1C. An inbuilt assumption of the 1D reduced-dimension model is that $\left(u_{i,J-1}^{\tau} - 2u_{i,J}^{\tau} + u_{i,J+1}^{\tau}\right) = 0$, the number of molecules leaving the focal plane, $-2u_{i,J}^{\tau}$, is equal to the number of molecules entering it, $u_{i,J-1}^{\tau} + u_{i,J+1}^{\tau}$, at all points in space, $i$, for all time, $\tau$: There is zero-flux through the focal plane.

**When is the zero-flux assumption valid?**

The zero-flux assumption is valid for a subset of reduced-dimension models, those for which a mesh can be drawn such that the zero-flux assumption holds. For example, consider the formation of a polarity patch on a spherical cell, one could construct a spherical mesh with a pole located at the centre of the patch, Fig. 1A. Because of the placing of the mesh concentrations $u_{i,J-1}^{\tau}$, $u_{i,J}^{\tau}$, and $u_{i,J+1}^{\tau}$ are equal and the zero-flux assumption holds, Fig. 1E. Thus, molecular movement in a 1D reduced-dimension model of this system would be accurately described by the 1D diffusion equation.

An example of a system for which the zero-flux assumption does not hold is the formation of a polarity patch along the body of an elongated cell, Fig. 1A. To solve the diffusion equation here a square mesh is constructed on the cell surface. As a polarity patch is radial and the mesh is square a focal plane cannot be found such that a 1D reduced-dimension model would obey the zero-flux assumption, Fig. 1D. Thus, the 1D diffusion equation would give an inaccurate approximation of 2D molecular movement through the focal plane.

For the modeller, the validity of the zero-flux assumption can be ascertained without generating a solution for the full system and calculating molecular flux through the focal plane. The modeller can consider the symmetries in the full system and the form of the full mesh using cartoons, as in Fig. 1. From the cartoons one can estimate whether or not the zero-flux assumption holds. This cartoon estimation of zero-flux is not dissimilar to the estimation made by the modeller when deciding whether or not it is appropriate to reduce a system's dimension.

**Calculating molecular flux through the focal plane**

Here we present a general methodology which can be used to increase the accuracy of reduced-dimension models that do not satisfy the zero-flux assumption: If it is possible to estimate the concentrations either side of the focal plane (terms $u_{i,J+1}^{\tau}$ and $u_{i,J-1}^{\tau}$ in the explicit 2D-FDM equation) using the concentrations on the focal plane ($u_{i,J}^{\tau}$), then we can estimate the flux through the focal plane. Reduced-dimension models are utilised in systems exhibiting symmetry, thus one can exploit the very symmetry used for dimension reduction to estimate concentrations on either side of the focal plane.

**Solving the 2D diffusion equation in a 1D reduced-dimension model**

We will use the general methodology to construct a FDM that solves the 2D diffusion equation in a 1D reduced dimension model. Consider again the polarity patch on the body of an elongated cell, Fig. 1A, D, F. The polarity patch has radial symmetry. A property of radial symmetry is that the concentration profile along all lines running through the centre of the patch is identical. In a 1D reduced-dimension model we reduce the dimension such that the 1D model calculates the concentration profile along one line running through the centre of the 2D patch, (row $J$). As the concentration profile on all lines running through the centre of the patch is identical we use the concentration profile in the 1D model to calculate the concentrations at all points on the 2D surface. To calculate the molecular flux through the focal plane, row $J$, we need only estimate concentrations on either side of the focal plane, rows $J \pm 1$. Interpolation on the 1D concentration profile is used to estimate the concentrations either side of the focal plane. The interpolation mesh points are found using Pythagoras' theorem Fig. 1F-G, Methods M1. In order to use Pythagoras to calculate the interpolation mesh points, the modeller has to set a value for $\Delta y$. In the illustrative examples below, we have set $\Delta y = \Delta x$. Analysis on the accuracy of estimating concentrations either side of the focal plane and the choice of $\Delta x$, $\Delta y$, can be found in SI Appendix 1. Further discussion on the derivation of the explicit 1D finite difference diffusion equation with unrestricted movement (explicit 1D-uFDM) can be found in Methods M2. The 1D-uFDM is,

$$u_i^{\tau+1} = u_i^\tau + \frac{\Delta t}{\Delta x^2} D(u_{i-1}^\tau - 2u_i^\tau + u_{i+1}^\tau) + \frac{\Delta t}{\Delta y^2} D\left(\tilde{u}_{i,J-1}^\tau - 2u_i^\tau + \tilde{u}_{i,J+1}^\tau\right) \qquad \text{explicit 1D-uFDM}$$

where $\tilde{u}_{i,J\pm1}^\tau$ denotes the estimated concentrations on rows $J \pm 1$. See Methods M3 for the solution to the explicit 1D-uFDM. The explicit 1D-uFDM numerically stability condition is derived in Methods M4 and tested numerically in SI Appendix 3. A fully implicit 1D-uFDM is ill defined, SI Appendix 2, however a semi-implicit 1D-uFDM and numerical stability condition can be derived, Methods M5-7. The explicit and semi-implicit 1D-uFDMs solve the 2D diffusion equation in a 1D reduced-dimension model.

In all three illustrative examples below, we are testing the ability of the explicit 1D-uFDM to solve the 2D diffusion equation in a 1D reduced-dimension model. We will also illustrate the accuracy gained by solving the 2D diffusion equation in a 1D reduced-dimension model, when compared to solving the 1D diffusion equation. Thus, the solutions of the 1D-uFDM and 1D-FDM will be compared to the solution on a slice through the centre of the patch in the full system. The 2D solution will be calculated using 2D-FDM.

**Illustrative example 1: Diffusion**

First, we considered the diffusion of molecules, $u$, from a concentrated patch on the membrane. The initial concentration profile in the reduced dimension models (Fig. 2A) was identical to the initial concentration along a slice through the centre of the patch in the 2D system (Fig. 2A, inset). The diffusion equation was solved using the explicit 2D-FDM, 1D-FDM and 1D-uFDM, diffusion coefficient $D = 0.1\ \mu m^2 s^{-1}$, Methods M8. The solutions of $u$ in both 1D reduced-dimension models were compared with the solutions of $u$ along a slice through the centre of the initial patch (see SI Appendix 8 for implicit 2D-FDM, 1D-FDM and semi-implicit 1D-uFDM comparisons). Results: The reduced-dimension model using the 1D diffusion equation to describe diffusive movement was quantitatively inaccurate at estimating the concentration on a slice through the centre of the 2D patch, Fig. 2B-C. This is because on the 2D membrane molecules diffuse out of the focal plane resulting in a reduction in the mean concentration of molecules in the central slice, Fig. 2D. However, molecules in the 1D-FDM model are trapped in the focal plane and the mean concentration of molecules remains constant, resulting in a higher homogeneous steady state, Fig. 2B-D. The 1D-uFDM estimates the flux out of the focal plane producing a more accurate reduced-dimension representation of the full system, Fig. 2B-E. While the 1D-uFDM represents the 2D system well it does contain error, Fig. 2E, in depth error analysis can be found in SI Appendix 6 and 7.

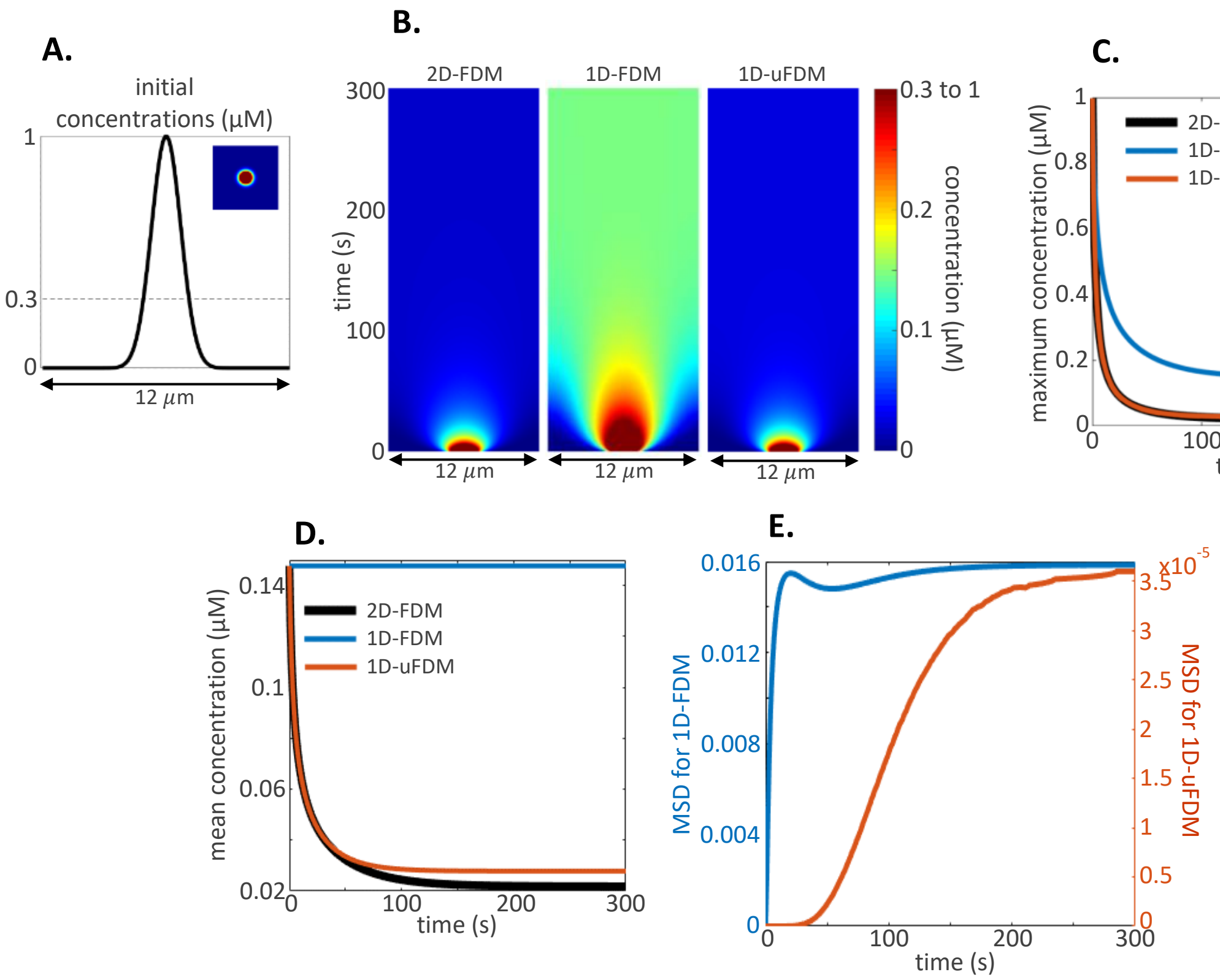

**Figure 2. (A)** Initial concentrations for diffusion solutions, $u_{1D}(x,0) = e^{-x^2}$ and inset $u_{2D}(x,y,0) = e^{-(x^2+y^2)}$. The dotted line shows the threshold for colorplots (inset and (B)). The threshold was chosen to highlight differences in dynamics and homogeneous steady states in kymographs. **(B)** Kymographs of the 2D solution along a slice through the centre of the initial patch, and the 1D solutions. **(C)** Concentration at the centre of the initial patch plotted over time for each solution. In this illustrative example the central concentration is the maximum concentration. **(D)** Mean concentration in the 1D solutions compared with the mean concentration of the 2D solution along a slice through the centre of the initial patch. **(E)** Mean squared distance (MSD) between the concentrations along a slice through the centre of the initial patch in the 2D solution and the 1D solutions. 1D-FDM comparison shown in blue, 1D-uFDM comparison shown in red.

**Illustrative example 2: Florescence recovery after photobleaching**

The 1D-uFDM was able to solve the 2D diffusion equation in a 1D reduced-dimension model when calculating molecules diffusing away from a central patch. We went on to ask, how accurate is the 1D-uFDM when calculating the movement of molecules into a central trough. To answer this question we modelled a florescence recovery after photobleaching (FRAP) experiment. FRAP experiments are used to estimate the diffusion coefficient of a fluorescently tagged protein. In a FRAP experiment fluorophores, attached to a protein of interest, are bleached by a laser. The florescence recovery within the bleached area is recorded and used to estimate a diffusion coefficient.

In our FRAP model the initial fluorophore concentrations were set to reflect a uniformly covered membrane after bleaching with a Guassian laser [18], Fig. 3A. Diffusive movement of the unbleached, fluorescently tagged, proteins was modelled using the 2D-FDM, the 1D-FDM and the 1D-uFDM, diffusion coefficient $D = 0.1\ \mu m^2 s^{-1}$. Results: In all FRAP solutions fluorescently tagged proteins diffused into the bleached area, Fig. 3B. The 1D-FDM solution had a low homogeneous steady state concentration when compared to the homogeneous steady state along a slice through the centre of the bleached area in the full system, Fig. 3B. The low steady state concentration in the 1D-FDM solution can be accounted for by the inability of tagged proteins to move into the focal plane, Fig. 3C. The 1D-uFDM estimated the movement of tagged proteins into the focal plane to give a more accurate representation of the 2D system Fig. 3B, C. Further accuracy analysis can be found in SI Appendix 9.

As FRAP is used to estimate diffusion coefficients we performed FRAP analysis on our simulated FRAP data, Methods M9. The FRAP recovery curves, Fig. 3D, show the mean florescence recovery within the region of interest (ROI, dotted lines Fig. 3A). The results of FRAP analysis can be found in Fig. 3E. The half time, $t_{1/2}$, is the time needed for the florescence intensity to reach half its maximum recovery. For the 2D system we calculated the half time and estimated the diffusion coefficient, $\tilde{D}$, using the full 2D solution and the data along a slice through the centre of the bleached area. Both methods provided comparable results, estimating the diffusion coefficient accurately to one decimal place, Fig. 3E. FRAP analysis on the 1D-FDM estimated the diffusion coefficient to be half its actual value. In order for a 1D-FDM solution to estimate a diffusion coefficient $\tilde{D} = 0.1\ \mu m^2 s^{-1}$, the actual diffusion coefficient had to be increased to $D = 0.22\ \mu m^2 s^{-1}$, Fig. 3E. FRAP analysis on the 1D-uFDM solution resulted in an accurate estimation of the diffusion coefficient. Thus, the 1D-uFDM is able to estimate 2D diffusive movement into a trough. Furthermore, these results show that data obtained from membrane FRAP experiments could be used to fit parameters in 1D-uFDM reduced-dimension models.

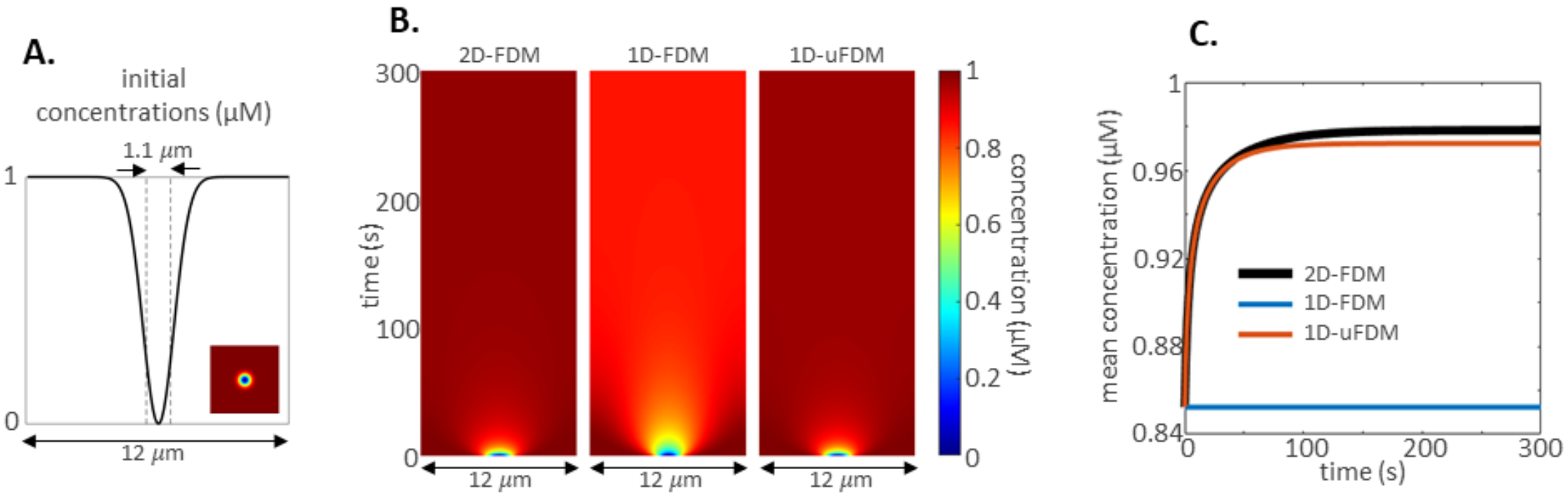

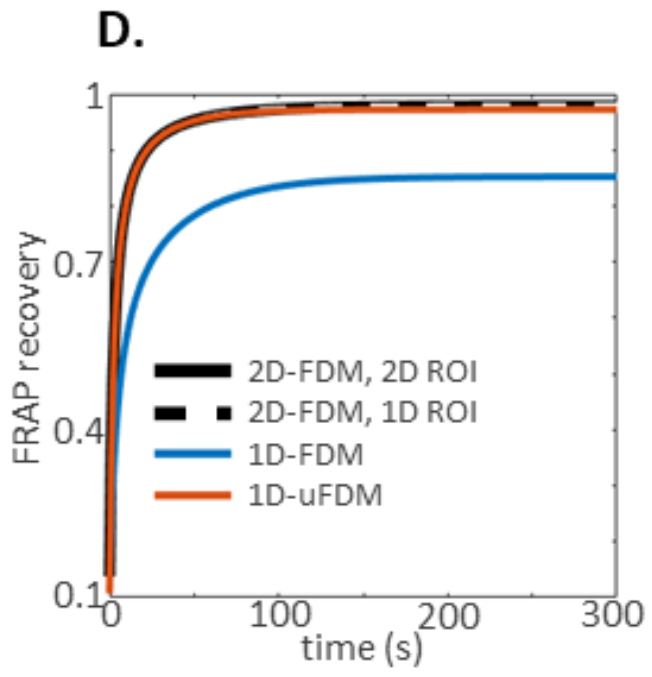

E.

| $D = 0.1\ \mu m^2 s^{-1}$ | $t_{1/2}\ s$ | $\tilde{D}$ |
|---|---|---|
| 2D-FDM, 2D ROI | 2.711 | 0.106 |
| 2D-FDM, 1D ROI | 2.801 | 0.103 |
| 1D-FDM | 6.145 | 0.047 |
| 1D-uFDM | 2.678 | 0.108 |
| $D = 0.22\ \mu m^2 s^{-1}$ | | |
| 1D-FDM | 2.678 | 0.108 |

**Figure 3. (A)** Initial concentrations for FRAP simulations after bleaching with a Guassian laser, radius $0.55\ \mu m$ [18]. $u_{1D}(x, 0) = 1 - e^{-x^2}$, inset $u_{2D}(x, y, 0) = 1 - e^{-(x^2+y^2)}$. Dotted lines show the ROI. **(B)** Kymographs of the 2D solution along a slice through the centre of the bleached area, and the 1D solutions. **(C)** Mean concentration in the 1D solutions compared with the mean concentration of the 2D solution along a slice through the centre of the bleached area. **(D)** FRAP recovery curves. The 2D ROI is a circle diameter $1.1\ \mu m$, 1D ROI a line length $1.1\ \mu m$. **(E)** Results

of FRAP analysis on the 1 and 2D solutions. Half time, $t_{1/2}$, actual diffusion coefficients, $D$, and estimated diffusion coefficients, $\widetilde{D}$.

**Illustrative example 3: Reaction diffusion dynamics**

Spatial models rarely focus solely on diffusion. We asked to what extent could 1D-uFDM reaction-diffusion (RD) model capture 2D RD dynamics along a slide through the full system's symmetry. To address this question, a two component, mass conserved, substrate depletion model was used [8], Fig. 4A, Methods M10,

$$\frac{\partial}{\partial t}u_1 = u_1^2 u_2 - u_1 + (0.01)\nabla^2 u_1$$
$$\frac{\partial}{\partial t}u_2 = -(u_1^2 u_2 - u_1) + \nabla^2 u_2$$

$\nabla^2$ denotes diffusion, the Laplacian. Initial concentrations were set to represent a signalling event which caused a pulse conversion of molecules to $u_1$ from the $u_2$ pool, Fig. 4B. After this initial signalling event the dynamics of the system were described by the RD equations. As in previous illustrative examples, the RD equations were solved in the full 2D system and the two 1D reduced-dimension models. Diffusion was solved using the 2D-FDM, the 1D-FDM and the 1D-uFDM. The RD model was used to explore the effect of a progressively stronger initial signalling pulse on the dynamics of each solution. Results: For all solutions, larger initial signalling pulses deplete local $u_2$ such that the RD positive feedback becomes ineffective ($u_1^2 u_2$ is very small) and a $u_1$ trough is soon formed at the location of the initial pulse, Fig. 4C, E. Similar to the FRAP analysis results, $u_1$ molecules were replenished more slowly in the 1D-FDM RD solution than in the 2D-FDM and 1D-uFDM RD solutions. Thus, in the 1D-FDM RD solution, as the initial signalling pulse increases, creating a larger trough, movement into the trough becomes insufficient to reach the centre and two narrow peaks are formed, Fig. 4C-D. The 1D-FDM solution is qualitatively different from the 2D-FDM solution for $\alpha \geq 1.5$, Fig. 4C-E. The 1D-uFDM reproduced all the 2D-FDM RD molecular dynamics through the focal plane, Fig. 4C-E, showing that it can be used to increase the accuracy of reduced-dimension RD models. Further results and accuracy analysis can be found in SI Appendix 10.

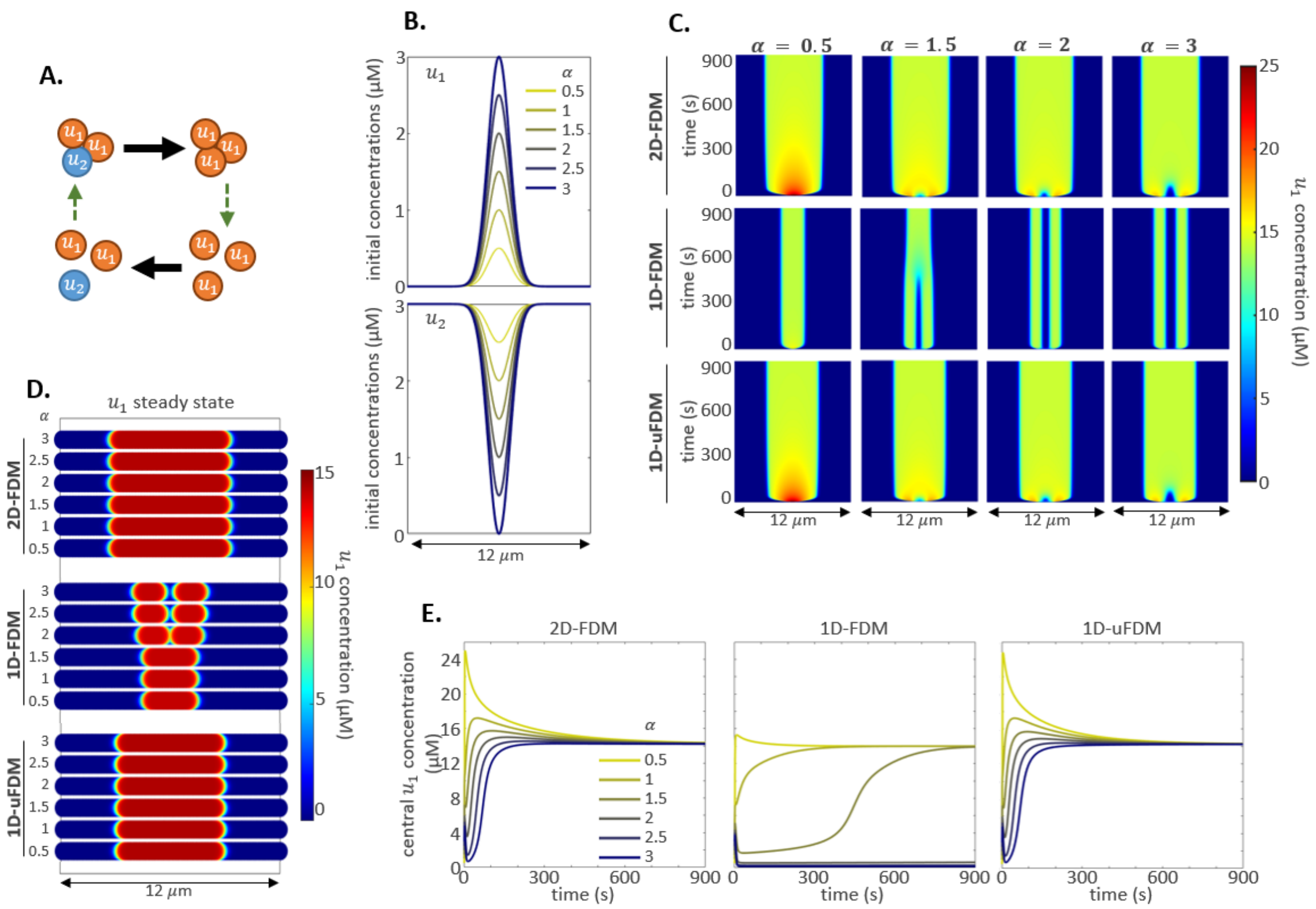

**Figure 4. (A)** Cartoon of the RD model. Black arrows represent reactions, green dashed arrows represent diffusion. Mass is conserved, molecules are only changed, never created or lost. **(B)** Initial concentrations of $u_1$ and $u_2$ for the RD models. $\alpha = [0.5, 1, \ldots, 3]$ is the maximum concentration of the $u_1$ peak after the initial signalling pulse. $u_1(x, y, 0) = \alpha e^{-(x^2+y^2)}$, $u_1(x, 0) = \alpha e^{-x^2}$, and $u_2 = m - u_1$, where $m$ is the mass of the system, see Methods M10. **(C)** Kymographs of the $u_1$ concentration in the 2D solution along a slice through the centre of the bleached area, and the 1D solutions. **(D)** Steady state colorplots of the 2D and 1D $u_1$ solutions. **(E)** $u_1$ concentration dynamics at the centre of the initial peak in the 1D and 2D solutions.

## Methods

### M1: Generating the interpolation mesh in 1D

Consider a 1D reduced-dimension model, reduced from a 2D uniform mesh (Fig. 1A). To solve the 1D-uFDM one must use the concentrations on the 1D mesh, row $J$ of the 2D mesh, to estimate concentrations at mesh-points in rows $J \pm 1$ of the 2D mesh (Fig. 1D, F and G). The reduced-dimension model is a ring and so has periodic boundary conditions. Without loss of generality, we will assume the centre of the patch at $i = N/2$. To generate the mesh to be interpolated from re-index the 1D mesh using $n = i - N/2$. The transformed index, to be interpolated from is $n = \left\{\left(1 - \frac{N}{2}\right), \left(2 - \frac{N}{2}\right), \ldots - 1, 0, 1, \ldots \left(\frac{N}{2} - 1\right), \frac{N}{2}\right\}$. To generate the mesh to be interpolated to use the transformed index to calculate the distance of each point in rows $J \pm 1$ from the centre of the patch using the equation $\sqrt{(\mathrm{n}\Delta\mathrm{x})^2 + \Delta\mathrm{y}^2}$ for $n = \left\{\left(1 - \frac{N}{2}\right), \left(2 - \frac{N}{2}\right), \ldots - 1, 0, 1, \ldots \left(\frac{N}{2} - 1\right), \frac{N}{2}\right\}$, Fig. 1G. Then interpolate. For systems with radial symmetry one need only estimate concentrations at a quarter of the phantom points to solve the 1D-uFDM, for example, $n = \left\{0, 1, \ldots \left(\frac{N}{2} - 1\right), \frac{N}{2}\right\}$. The accuracy of estimating concentrations in rows $J \pm 1$ of the 2D mesh, using interpolation, is discussed in SI Appendix 1.

**M2: Explicit 1D-uFDM: derivation**

The explicit 2D-FDM is,

$$u_{i,j}^{\tau+1} = u_{i,j}^{\tau} + d_x\left(u_{i-1,j}^{\tau} - 2u_{i,j}^{\tau} + u_{i+1,j}^{\tau}\right) + d_y\left(u_{i,j-1}^{\tau} - 2u_{i,j}^{\tau} + u_{i,j+1}^{\tau}\right)$$

where $d_x = \frac{\Delta t}{\Delta x^2}D$ and $d_y = \frac{\Delta t}{\Delta y^2}D$. $J$ is the row of mesh points through the centre of the patch, the focal plane. When estimating 2D diffusion, through the centre of the patch, in 1D space, we only have information about row $J$. Thus, we have to estimate the concentrations $u_{i,J-1}^{\tau}$ and $u_{i,J+1}^{\tau}$. To achieve this we use the property of radial symmetry exhibited by a patch of proteins. Transform the $\Delta x$ mesh points such that the centre of the patch is at mesh point $n = 0$ (see previous Methods M1). The property of radial symmetry dictates that, for $k \geq 0$, $u_{k,J+1}^{\tau} = u_{k,J-1}^{\tau} = u_{-k,J+1}^{\tau} = u_{-k,J-1}^{\tau}$ (Fig. 1D and F). Denote the estimated concentration $u_{n,J\pm1}^{\tau}$ as $u_{\sqrt{(n\Delta x)^2+\Delta y^2}}^{\tau}$ $\forall\, n = \left\{\left(1-\frac{N}{2}\right), \dots \frac{N}{2}\right\}$ (Fig. 1G). Substituting the interpolated values of $u_{n,J\pm1}^{\tau}$ into the 2D-FDM equations, and removing the $J$ subscript we get the 1D-uFDM,

$$u_n^{\tau+1} = u_n^{\tau} + d_x(u_{n-1}^{\tau} - 2u_n^{\tau} + u_{n+1}^{\tau}) + 2d_y\left(u_{\sqrt{(n\Delta x)^2+\Delta y^2}}^{\tau} - u_n^{\tau}\right)$$

**M3: Explicit 1D-uFDM: solution**

To solve the explicit 1D-uFDM we write the 1D-uFDM equation in matrix form,

$$\underline{u}^{\tau+1} = A\underline{u}^{\tau} + 2d_y\underline{\tilde{u}}_J^{\tau}$$

Where, $\underline{u}^{\tau}$ denotes the $N$x1 vector of concentrations $u_n^{\tau}$ on mesh points $n = \left\{\left(1-\frac{N}{2}\right), \dots \frac{N}{2}\right\}$, $\underline{\tilde{u}}_J^{\tau}$ denotes the $N$x1 vector of interpolated concentrations $u_{\sqrt{(n\Delta x)^2+\Delta y^2}}^{\tau}$, $n = \left\{\left(1-\frac{N}{2}\right), \dots, \frac{N}{2}\right\}$, and

$$A = \begin{bmatrix} 1-2(d_x+d_y) & d_x & & & d_x \\ d_x & 1-2(d_x+d_y) & d_x & & \\ & & \ddots & & \\ & & d_x & 1-2(d_x+d_y) & d_x \\ d_x & & & d_x & 1-2(d_x+d_y) \end{bmatrix}$$

a tridiagonal $N$x$N$ matrix, with periodic boundary conditions. Using interpolation on the matrix form we derive the solution to the explicit 1D-uFDM,

$$\underline{u}^{\tau} = \underline{\lambda}^{\tau}\underline{u}^{0} + 2d_y\sum_{k=1}^{\tau}\underline{\lambda}^{\tau-k}\underline{\tilde{u}}_J^{k-1}$$

Where $\underline{\lambda}$ is the vector of eigenvalues for $A$.

**M4: Explicit 1D-uFDM: numerical stability condition**

To calculate the stability condition for 1D-uFDM numerical stability recall that the values $\underline{\tilde{u}}_J^{\tau}$ are interpolated from $\underline{u}^{\tau}$ at time $\tau$, thus, as long as a stable interpolation method is used, the solution will be stable if $|\lambda| \leq 1\ \forall\, \lambda \in \underline{\lambda}$. Gerschgorin's circle theorem [17] states that $\lambda$ is bounded by the inequality,

$$1 - 2(d_x + d_y) - 2d_x \leq \lambda \leq 1 - 2(d_x + d_y) + 2d_x$$

Which can be simplified to,

$$1 - 4d_x - 2d_y \leq \lambda \leq 1 - 2d_y$$

Thus, the 1D-uFDM solution will be numerically stable if, $1 - 2d_y \leq 1$ and $-1 \leq 1 - 4d_x - 2d_y$. $1 - 2d_y \leq 1$ is always satisfied as $d_y > 0$. The inequality, $-1 \leq 1 - 4d_x - 2d_y$ leads to the explicit 1D-uFDM stability condition,

$$2d_x + d_y \leq 1$$

The explicit 1D-uFDM stability condition is numerically verified in SI Appendix 3.

**M5: Semi-implicit 1D-uFDM: derivation**

A fully implicit 1D-uFDM is ill defined as flux through the focal plane is inferred using the concentrations on the focal plane, SI Appendix 2. A semi-implicit numerical solver can be defined in which flux through the focal plane is solved explicitly and flux on the focal plane is solved implicitly. The semi-implicit 2D-FDM equation is,

$$u_{i,j}^{\tau+1} = u_{i,j}^{\tau} + d_x\left(u_{i-1,j}^{\tau+1} - 2u_{i,j}^{\tau+1} + u_{i+1,j}^{\tau+1}\right) + d_y\left(u_{i,j-1}^{\tau} - 2u_{i,j}^{\tau} + u_{i,j+1}^{\tau}\right)$$

Using the same reasoning used for the derivation of the explicit 1D-uFDM, the semi-implicit 1D-uFDM is,

$$u_n^{\tau+1} = u_n^{\tau} + d_x(u_{n-1}^{\tau+1} - 2u_n^{\tau+1} + u_{n+1}^{\tau+1}) + 2d_y\left(u^{\tau}_{\sqrt{(n\Delta x)^2+\Delta y^2}} - u_n^{\tau}\right)$$

Note, the semi-implicit 1D-uFDM scheme mirrors the 1D Crank-Nicolson method which converges and is unconditionally stable[17]. We will show that the same is not true for the semi-implicit 1D-uFDM. However, the semi-implicit 1D-uFDM has less strict numerical stability conditions than the explicit 1D-uFDM.

**M6: Semi-implicit 1D-uFDM: solution**

To solve the semi-implicit 1D-uFDM we write it in matrix form,

$$\underline{u}^{\tau+1} = C^{-1}\left(\left(1-2d_y\right)\underline{u}^{\tau} + 2d_y\underline{\tilde{u}}_J^{\tau}\right)$$

where,

$$C = \begin{bmatrix} 1+2d_x & -d_x & & & -d_x \\ -d_x & 1+2d_x & -d_x & & \\ & & \ddots & & \\ & & -d_x & 1+2d_x & -d_x \\ -d_x & & & -d_x & 1+2d_x \end{bmatrix}$$

a tridiagonal $N$x$N$ matrix, with periodic boundary conditions. Again, using interpolation on the matrix form of the equation we derive the solution to the semi-implicit 1D-uFDM,

$$\underline{u}^{k+1} = \frac{1}{\lambda^{k+1}}\left(1-2d_y\right)^{k+1}\underline{u}^0 + 2d_y\sum_{l=1}^{k}\left(1-2d_y\right)^{l-1}\frac{1}{\lambda^l}\underline{\tilde{u}}_J^{k-l}$$

Where $\underline{\lambda}$ is the vector of eigenvalues for $C$.

**M7: Semi-implicit 1D-uFDM: numerical stability condition**

For the solution to the semi-implicit 1D-uFDM to be stable two inequalities must hold, $|1/\lambda| \leq 1$, and $\left|1-2d_y\right| \leq 1$. The first inequality $1 \leq |\lambda|$ can be investigated using Gerschgorin's circle theorem [17]. Gerschgorin's circle theorem states that,

$$1 + 2d_x - 2d_x \leq \lambda \leq 1 + 2d_x + 2d_x$$

Which can be simplified to,

$$1 \leq \lambda \leq 1 + 4d_x$$

Thus, the inequality $1 \leq |\lambda|$ is always satisfied. The second inequality $\left|1-2d_y\right| \leq 1$expands to, $-1 \leq 1-2d_y \leq 1$. $1-2d_y \leq 1$ as $d_y \geq 0$. $-1 \leq 1-2d_y$ leads to the semi-implicit 1D-uFDM stability condition,

$$d_y \leq 1$$

The semi-implicit 1D-uFDM stability condition is numerically verified in SI Appendix 3.

**M8: Illustrative example 1: Parameters**

For results shown in Fig. 2B to E, to enable comparisons between the different numerical method solutions, the explicit 2D-FDM, 1D-FDM and 1D-uFDM were all solved using the same parameters: $D = 0.1\ \mu m^2 s^{-1}$, $\Delta x = \Delta y = 0.1\ \mu m$, $\Delta t = 0.01\ s$. These values were chosen to satisfy the numerical stability conditions of all three numerical methods. The explicit 1D-uFDM accuracy analysis found in SI Appendix 6 and 7 was also taken into consideration.

**M9: Illustrative example 2: Parameters and FRAP analysis**

For the comparative FRAP analysis, $D = 0.1\ \mu m^2 s^{-1}$, $\Delta x = \Delta y = 0.1\ \mu m$, $\Delta t = 0.01\ s$ for all solutions. The 2D ROI used to calculate the FRAP recovery curve was a circle radius $0.55 \mu m$ and the 1D ROI a line $1.1 \mu m$ in length (Fig. 3A). Both ROIs were placed in the centre of the bleached region. $t_{1/2}$ was calculated using linear interpolation on the FRAP recovery curve, Fig. 3D. To estimate the diffusion coefficient the equation,

$$\widetilde{D} = \frac{r_n^2 + r_e^2}{8 t_{1/2}}$$

was used[18], where $r_e$ represents the value of the effective radius and $r_n$ the laser radius. For the FRAP simulations, $r_e = \sqrt{2}$ and $r_n = 0.55$.

**M10: Illustrative example 3: RD parameters and simulations**

Symmetry breaking parameters, and a mass sufficient for the system to exhibit saturation due to substrate depletion ($m = 3$) were chosen [8]. The mass of the system was defined as the mean concentration of molecules,

$$m = \langle u_1 \rangle + \langle u_2 \rangle$$

where, for $k = \{1,2\}$,

$$\langle u_k \rangle = \frac{1}{NM} \sum_{i=1}^{N} \sum_{j=1}^{M} u_{k(i,j)}$$

for the 2D case, and,

$$\langle u_k \rangle = \frac{1}{N} \sum_{i=1}^{N} u_{k(i)}$$

for 1D.

$\Delta x = \Delta y = 0.1\ \mu m$. Solving the RD equations was a done using a two-step process: reactions were solved using Euler's method at time steps $\Delta t_R = 1\text{x}10^{-5}\ s$ for all simulations, diffusion was solved using explicit 2D-FDM, 1D-FDM and 1D-uFDM at time steps $\Delta t_D = 0.002\ s$ to ensure numerical stability of the explicit 2D-FDM.

## Acknowledgments

This work was funded by the BBSRC, Project Grant BB/N009339/1.

## Supplementary Information

**Appendix 1: Accuracy of estimating concentrations at phantom points using interpolation.**

Numerical solutions within this Brief Communication were performed in MATLAB. Spline interpolation was found to be the most accurate interpolation method in MATLAB for the concentration profiles investigated here (data not shown). However, if a concentration profile approaches zero with a steep gradient spline interpolation can return a small negative concentration. Any negative concentrations resulting from MATLAB spline interpolation were set to zero.

To investigate the accuracy of estimating concentrations at phantom mesh-points (those in rows $J \pm 1$) using MATLAB spline interpolation, 2D concentration profiles were generated and the actual concentrations on rows $J \pm 1$ were compared with the estimated concentrations, which had been interpolated from the 1D concentration profile (Methods). To quantify interpolation accuracy the sum squared distance (SSD) between the interpolated and actual concentration profiles was calculated.

The concentration profiles used to investigate the accuracy of estimating concentrations at phantom points can be seen in Fig. S1. To better understand the influence of $\Delta x$ and $\Delta y$ on the estimation of phantom Point concentrations, a range of $\Delta x$ and $\Delta y$ values, between $0.001\ \mu m$ and $0.1\ \mu m$, was used for each concentration profile. As expected, increasing $\Delta x$ decreases the accuracy of the interpolation (Fig. S2, column 1). Comparing columns 1 and 2 of Fig. S2 we see that the error around the concentration peak, in the central 10% of the mesh, accounts for most of the interpolation error. Indeed, the greater the curvature of the concentration profile the greater the interpolation error. In order to estimate concentrations at phantom points on the special boundaries one must interpolate outside the data (Fig. S5b). Fig. S2, column 3, shows that, as we may expect, the greater $\Delta y$ the greater the error in estimation outside the boundary.

Supplementary Table 1 shows the maximum and minimum SSD when estimating the phantom points of all the concentration profiles shown in Fig. S1a, c, and e. The estimation error is small for small $\Delta x$ and $\Delta y$ values (Fig S2 and Table S1). Decreasing $\Delta x$ increases computation time, to reduce estimation error for larger $\Delta x$ a small $\Delta y$ should be chosen (Fig. S2). The choices of $\Delta x$ and $\Delta y$ should be chosen to satisfy the stability conditions of the numerical method (Methods).

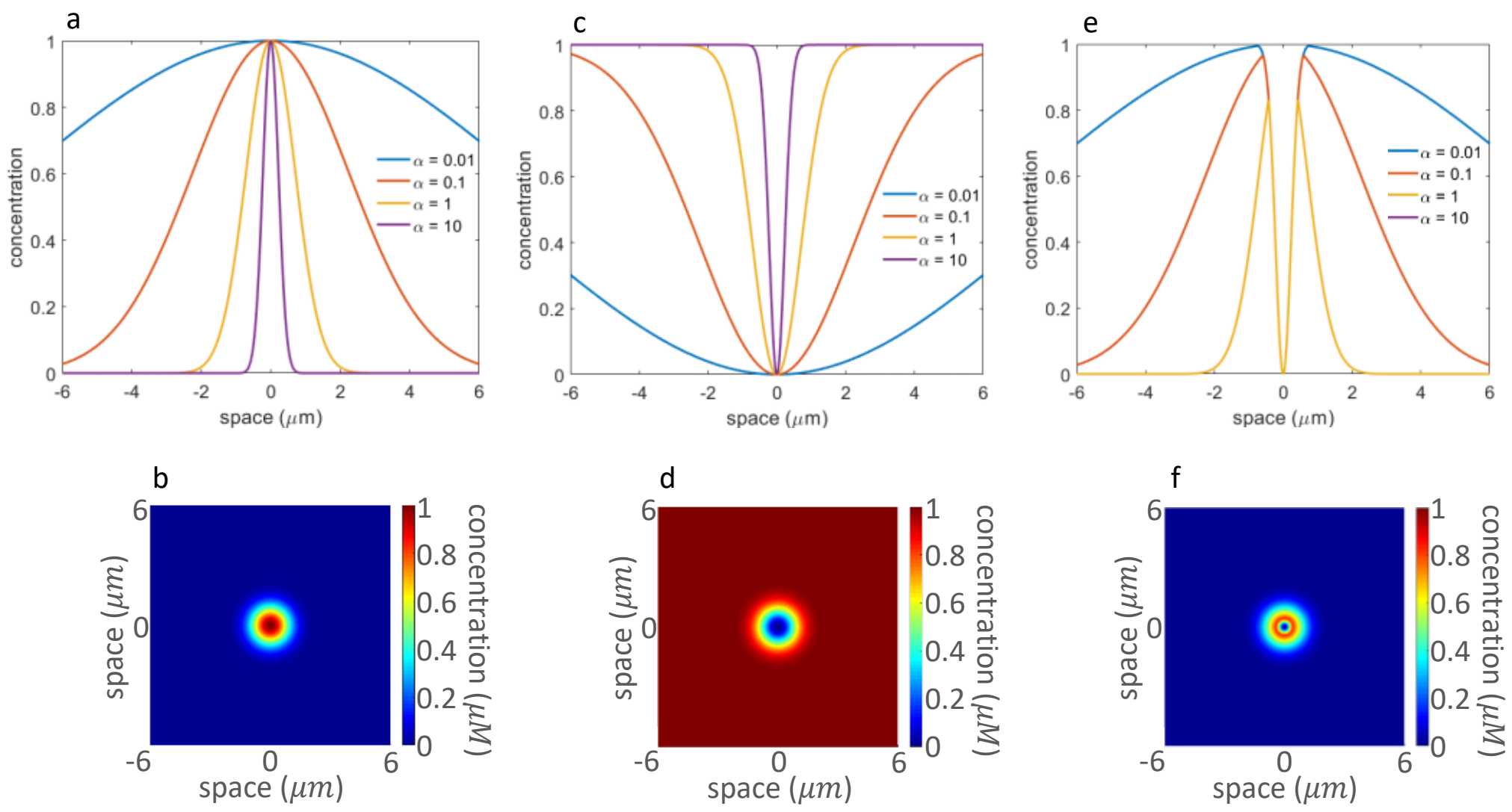

**Fig. S1: Concentration profiles used to investigate the accuracy of estimating concentrations at phantom points. (a)** 1D concentration profiles described by the equation, $\boldsymbol{u} = \boldsymbol{e}^{-\alpha x^2}$. **(b)** Example of the 2D system before dimension reduction reduced, $\boldsymbol{\alpha} = \mathbf{1}$, $\boldsymbol{u} = \boldsymbol{e}^{-(x^2+y^2)}$. **(c)** 1D concentration profiles described by the equation, $\boldsymbol{u} = \mathbf{1} - \boldsymbol{e}^{-\alpha x^2}$. **(d)** Example of the 2D system before dimension reduction, $\boldsymbol{\alpha} = \mathbf{1}$, $\boldsymbol{u} = \mathbf{1} - \boldsymbol{e}^{-(x^2+y^2)}$. **(e)** 1D concentration profiles described by the equation, $\boldsymbol{u} = \mathbf{min}\left(\boldsymbol{e}^{-\alpha_1 x^2}, \mathbf{1} - \boldsymbol{e}^{-\alpha_2 x^2}\right)$. $\boldsymbol{\alpha_2} = \mathbf{10}$ for all curves. $\boldsymbol{\alpha_1} = \mathbf{1}, \mathbf{0.1}, \mathbf{0.01}$ for the yellow, red, blue curve, respectively. **(f)** Example of the 2D system before dimension reduction, $\boldsymbol{\alpha_1} = \mathbf{1}$, $\boldsymbol{u} = \mathbf{min}\left(\boldsymbol{e}^{-(x^2+y^2)}, \mathbf{1} - \boldsymbol{e}^{-10(x^2+y^2)}\right)$.

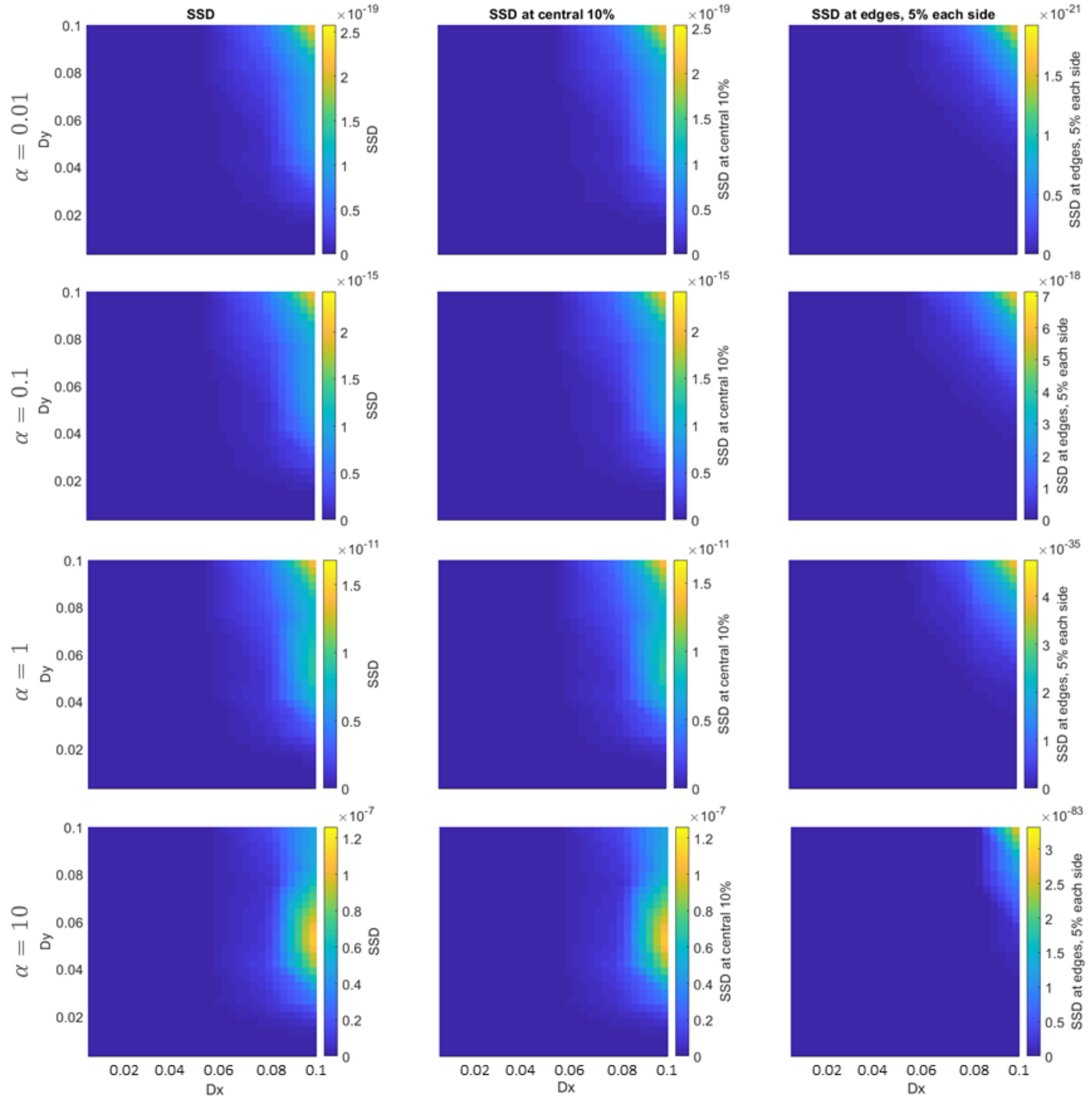

**Fig. S2: SSD between estimated concentrations at phantom points and actual concentrations in the 2D System.** The analysis was performed on the profiles, $\boldsymbol{u_{1D}} = \boldsymbol{e}^{-\alpha x^2}$ and $\boldsymbol{u_{2D}} = \boldsymbol{e}^{-\alpha(x^2+y^2)}$ (Fig. S1a and b).

**Table S1: Minimum and maximum SSDs between estimated concentrations at phantom points and actual concentrations in the 2D System.** Profiles analyzed are those in Fig. S1a, c and e. Minimum SSDs shown in blue text, maximum in black.

| Supplementary Figure | 1a | | | 1b | | | 1c | | | | |
|---|---|---|---|---|---|---|---|---|---|---|---|
| $\alpha$ | whole | central | edges | whole | centre | edges | $\alpha_1$ | $\alpha_2$ | whole | centre | edges |
| 0.01 | $9x10^{-30}$<br>$3x10^{-19}$ | $1x10^{-30}$<br>$3x10^{-19}$ | $3x10^{-31}$<br>$2x10^{-21}$ | $7x10^{-30}$<br>$3x10^{-19}$ | $8x10^{-31}$<br>$3x10^{-19}$ | $2x10^{-31}$<br>$2x10^{-21}$ | 10 | 0.01 | $6x10^{-16}$<br>$1x10^{-7}$ | $3x10^{-23}$<br>$1x10^{-7}$ | $3x10^{-31}$<br>$2x10^{-21}$ |
| 0.1 | $3x10^{-29}$<br>$2x10^{-15}$ | $4x10^{-30}$<br>$2x10^{-15}$ | $5x10^{-32}$<br>$7x10^{-18}$ | $5x10^{-29}$<br>$2x10^{-15}$ | $3x10^{-30}$<br>$2x10^{-15}$ | $4x10^{-31}$<br>$7x10^{-18}$ | 10 | 0.1 | $3x10^{-14}$<br>$6x10^{-6}$ | $3x10^{-14}$<br>$6x10^{-6}$ | $5x10^{-32}$<br>$7x10^{-18}$ |
| 1 | $3x10^{-27}$<br>$2x10^{-11}$ | $3x10^{-27}$<br>$2x10^{-11}$ | $2x10^{-53}$<br>$5x10^{-35}$ | $3x10^{-27}$<br>$2x10^{-11}$ | $3x10^{-27}$<br>$2x10^{-11}$ | 0<br>$3x10^{-30}$ | 10 | 1 | $9x10^{-13}$<br>$2x10^{-4}$ | $9x10^{-13}$<br>$2x10^{-4}$ | $2x10^{-53}$<br>$1x10^{-35}$ |
| 10 | $3x10^{-23}$<br>$1x10^{-7}$ | $3x10^{-23}$<br>$1x10^{-7}$ | $2x10^{-275}$<br>$3x10^{-83}$ | $3x10^{-23}$<br>$1x10^{-7}$ | $3x10^{-23}$<br>$1x10^{-7}$ | 0<br>0 | | | | | |

**Appendix 2: An implicit 1D-uFDM is ill defined.**

The implicit 2D-FDM is,

$$u_{i,j}^{\tau+1} = u_{i,j}^{\tau} + \frac{\Delta t}{\Delta x^2} D\left(u_{i-1,j}^{\tau+1} - 2u_{i,j}^{\tau+1} + u_{i+1,j}^{\tau+1}\right) + \frac{\Delta t}{\Delta y^2} D\left(u_{i,j-1}^{\tau+1} - 2u_{i,j}^{\tau+1} + u_{i,j+1}^{\tau+1}\right)$$

Using the same reasoning as for the derivation of the explicit 1D-uFDM, the implicit 1D-uFDM is,

$$u_n^{\tau+1} = u_n^{\tau} + \frac{\Delta t}{\Delta x^2} D(u_{n-1}^{\tau+1} - 2u_n^{\tau+1} + u_{n+1}^{\tau+1}) + 2\frac{\Delta t}{\Delta y^2} D\left(u_{\sqrt{(n\Delta x)^2+\Delta y^2}}^{\tau+1} - u_n^{\tau+1}\right)$$

The vector form of the implicit 1D-uFDM is,

$$\underline{u}^{\tau} = B\underline{u}^{\tau+1} + 2d_y \underline{\tilde{u}}_J^{\tau+1}$$

$$B = \begin{bmatrix} 1+2(d_x+d_y) & -d_x & & & -d_x \\ -d_x & 1+2(d_x+d_y) & & & \\ & & \ddots & & \\ & & & 1+2(d_x+d_y) & -d_x \\ -d_x & & & -d_x & 1+2(d_x+d_y) \end{bmatrix}$$

a tridiagonal $N$x$N$ matrix, with periodic boundary conditions.

The solution to the implicit 1D-uFDM is,

$$\underline{u}^{\tau+1} = B^{-1}\left(\underline{u}^{\tau} - 2d_y \underline{\tilde{u}}_J^{\tau+1}\right)$$

However, the values of $\underline{u}^{\tau+1}$ needed to interpolate the values $\underline{\tilde{u}}_J^{\tau+1}$, and so $\underline{\tilde{u}}_J^{\tau+1}$ in the above equation are not defined.

**Appendix 3: Numerical verification of stability Conditions.**

The explicit and semi-implicit 1D-uFDM numerical stability conditions were tested numerically. Recall, $D\frac{\Delta t}{\Delta x^2} = d_x$ and $D\frac{\Delta t}{\Delta y^2} = d_y$. Consider first the explicit 1D-uFDM stability condition in terms of $\Delta x$, $\Delta y$, $\Delta t$ and $D$, $2D\frac{\Delta t}{\Delta x^2} + D\frac{\Delta t}{\Delta y^2} \le 1$. For $\beta \ge 0$ we would expect a numerical solution with $2D\frac{\Delta t}{\Delta x^2} + D\frac{\Delta t}{\Delta y^2} = 1 + \beta$ to be numerically unstable and a numerical solution with $2D\frac{\Delta t}{\Delta x^2} + D\frac{\Delta t}{\Delta y^2} = 1 - \beta$ to be numerically stable. To test the explicit 1D-uFDM stability condition $\Delta x$, $\Delta y$ and $D$ were set and $\Delta t_{\pm}$ was calculated using the formula,

$$2D\frac{\Delta t_{\pm}}{\Delta x^2} + D\frac{\Delta t_{\pm}}{\Delta y^2} = 1 \pm \beta$$

$$\Delta t_{\pm} = \frac{1 \pm \beta}{\mathrm{D}}\left(\frac{\Delta x^2 \Delta y^2}{2\Delta y^2 + \Delta x^2}\right)$$

Numerical solutions were run for $\Delta x = 0.01, 0.02, \ldots, 0.1\ \mu m$, $\Delta y = 0.01, 0.02, \ldots, 0.1\ \mu m$, $D = 0.1\ \mu m^2 s^{-1}$ and $\Delta t = \Delta t_{\pm}\ s$, $\beta = 1x10^{-4}, 1x10^{-3}, 1x10^{-2}$. The initial condition for each simulation was $u(x,t) = u(x,0) = e^{-x^2}$ for $x \in [0,12]\,\mu m$ (Fig. S1a, $\alpha = 1$). Numerical solutions were calculated for one hour, or until they crashed as a result of numerical instability. Fig. S3a shows the stability boundaries being tested. Table S2 shows the smallest values of $\beta$, out of those tested, for which numerical solutions with $\Delta t_+$ were numerically unstable and $\Delta t_-$ were numerically stable. For the values of $\Delta x, \Delta y$ in the blue cells of Table S2, $\Delta t_-$ simulations with $\beta = 1\text{x}10^{-4}, 1\text{x}10^{-3}$ were numerically unstable. To understand how far from the stability boundary the blue celled values in Table S2 were, the $\Delta t_\pm$ equation was rearranged such that $\Delta t_\pm = \Delta t \pm \beta \Delta t$, where $\Delta t = \frac{1}{\text{D}}\left(\frac{\Delta x^2 \Delta y^2}{2\Delta y^2 + \Delta x^2}\right)$. The values $\Delta t \pm (1\text{x}10^{-2})\Delta t$ are shown in Table S3.

The semi-implicit 1D-uFDM stability condition was tested numerically using the same methodology, with $\Delta t_\pm = \frac{1 \pm \beta}{\text{D}} \Delta y^2$ (Fig. S3b, Table S4). Simulations with $\Delta t_+$ were numerically unstable and $\Delta t_-$ were numerically stable for all $\Delta x$, $\Delta y$ simulated and $\beta = 1\text{x}10^{-4}$.

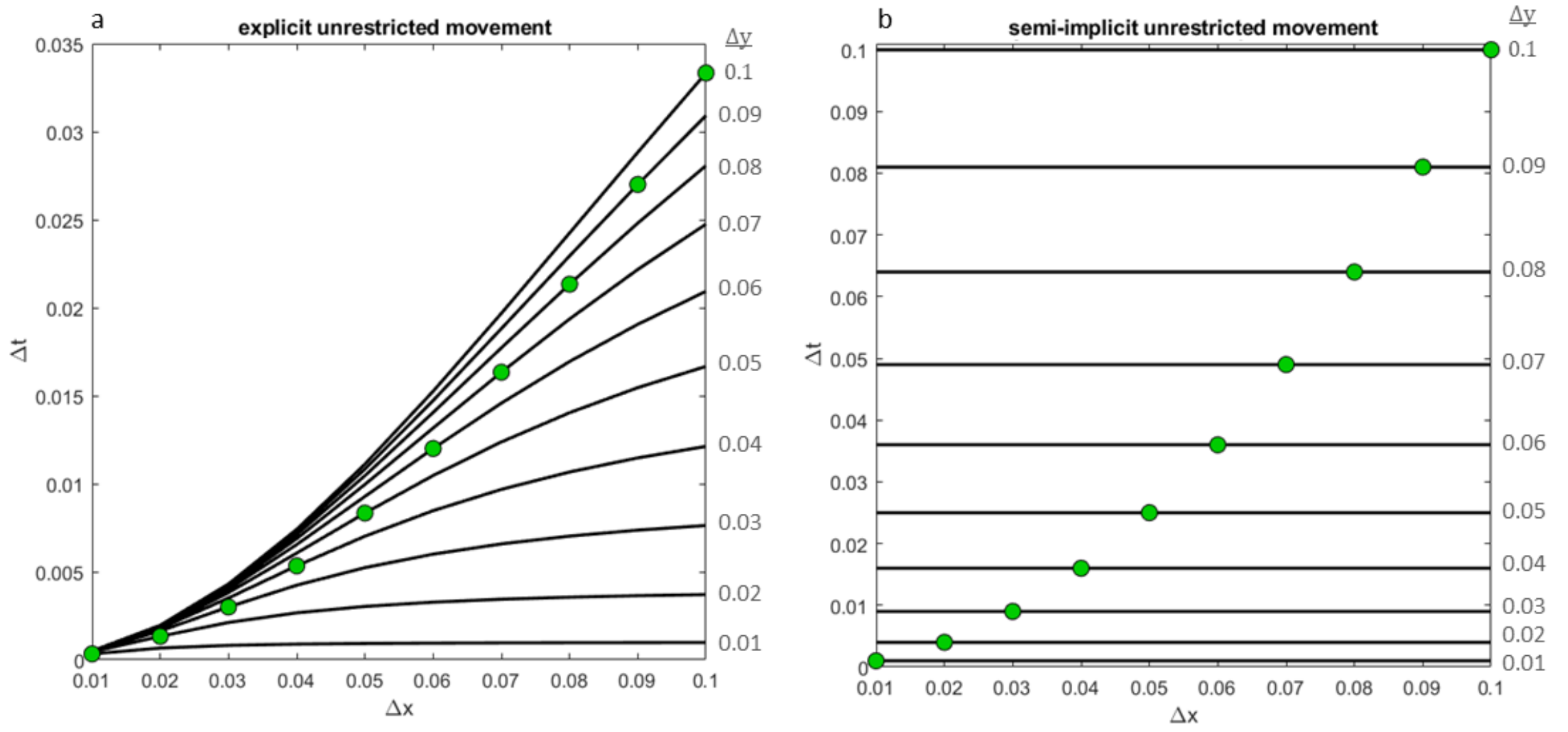

**Fig. S3: Numerical stability boundaries.** Black lines show the numerical stability boundaries for $\boldsymbol{\Delta y = 0.01, 0.02, \ldots, 0.1\ \mu m}$. $\boldsymbol{D = 0.1\ \mu m^2 s^{-1}}$ Values of $\boldsymbol{\Delta t}$ on or below the black lines are numerically stable. Green dots show the stability boundary when $\boldsymbol{\Delta x = \Delta y}$. Note, the $\boldsymbol{\Delta t}$ axis have different maxima on each graph. **(a)** Boundaries for numerical stability in the explicit 1D-uFDM are given by the equation, $\boldsymbol{\Delta t = \frac{1}{\text{D}}\left(\frac{\Delta x^2 \Delta y^2}{2\Delta y^2 + \Delta x^2}\right)\ s}$ . **(b)** Boundaries for numerical stability in the semi-implicit 1D-uFDM are given by the equation, $\boldsymbol{\Delta t = \frac{\Delta y^2}{D}\ s}$.

**Table S2: Results of the numerical test of the explicit 1D-uFDM stability condition.** Smallest values of $\boldsymbol{\beta}$ out of those tested ($\boldsymbol{\beta = 1\text{x}10^{-4}, 1\text{x}10^{-3}, 1\text{x}10^{-2}}$) for which simulations with $\boldsymbol{\Delta t_+}$ were unstable and $\boldsymbol{\Delta t_-}$ were stable. $\boldsymbol{D = 0.1\ \mu m^2 s^{-1}}$. Blue cells indicate the values of $\boldsymbol{\Delta x, \Delta y}$ for which $\boldsymbol{\Delta t_-}$ simulations with $\boldsymbol{\beta = 1\text{x}10^{-4}, 1\text{x}10^{-3}}$ were numerically unstable.

| | | $\Delta x$ | | | | | | | | | |
|---|---|---|---|---|---|---|---|---|---|---|---|
| | | **0.01** | **0.02** | **0.03** | **0.04** | **0.05** | **0.06** | **0.07** | **0.08** | **0.09** | **0.1** |
| $\Delta y$ | **0.01** | 0.0001 | 0.0001 | 0.0001 | 0.0001 | 0.0001 | 0.0001 | 0.0001 | 0.0001 | 0.0001 | 0.0001 |
| | **0.02** | 0.0001 | 0.0001 | 0.0001 | 0.0001 | 0.0001 | 0.0001 | 0.0001 | 0.0001 | 0.0001 | 0.0001 |
| | **0.03** | 0.01 | 0.0001 | 0.0001 | 0.0001 | 0.0001 | 0.0001 | 0.0001 | 0.0001 | 0.0001 | 0.0001 |
| | **0.04** | 0.01 | 0.0001 | 0.0001 | 0.0001 | 0.0001 | 0.0001 | 0.0001 | 0.0001 | 0.0001 | 0.0001 |
| | **0.05** | 0.01 | 0.0001 | 0.0001 | 0.0001 | 0.0001 | 0.0001 | 0.0001 | 0.0001 | 0.0001 | 0.0001 |

| | | | | | | | | | | |
|---|---|---|---|---|---|---|---|---|---|---|
| **0.06** | 0.01 | 0.01 | 0.0001 | 0.0001 | 0.0001 | 0.0001 | 0.0001 | 0.0001 | 0.0001 | 0.0001 |
| **0.07** | 0.01 | 0.01 | 0.0001 | 0.0001 | 0.0001 | 0.0001 | 0.0001 | 0.0001 | 0.0001 | 0.0001 |
| **0.08** | 0.01 | 0.01 | 0.0001 | 0.0001 | 0.0001 | 0.0001 | 0.0001 | 0.0001 | 0.0001 | 0.0001 |
| **0.09** | 0.01 | 0.01 | 0.01 | 0.0001 | 0.0001 | 0.0001 | 0.0001 | 0.0001 | 0.0001 | 0.0001 |
| **0.1** | 0.01 | 0.01 | 0.01 | 0.0001 | 0.0001 | 0.0001 | 0.0001 | 0.0001 | 0.0001 | 0.0001 |

**Table S3: Results of the numerical test of the explicit 1D-uFDM stability condition.** $\Delta t_{\pm} = \boldsymbol{\Delta t} \pm \boldsymbol{\beta \Delta t}$ for $\boldsymbol{\beta} = \mathbf{0.01}$.

| | | $\Delta x$ | | |
|---|---|---|---|---|
| | | **0.01** | **0.02** | **0.03** |
| $\Delta y$ | **0.01** | | | |
| | **0.02** | | | |
| | **0.03** | $4.7x10^{-4} \pm 4.7x10^{-6}$ | | |
| | **0.04** | $4.8x10^{-4} \pm 4.8x10^{-6}$ | | |
| | **0.05** | $4.9x10^{-4} \pm 4.9x10^{-6}$ | | |
| | **0.06** | $4.9x10^{-4} \pm 4.9x10^{-6}$ | $1.9x10^{-3} \pm 1.9x10^{-5}$ | |
| | **0.07** | $4.9x10^{-4} \pm 4.9x10^{-6}$ | $1.9x10^{-3} \pm 1.9x10^{-5}$ | |
| | **0.08** | $5.0x10^{-4} \pm 5.0x10^{-6}$ | $1.9x10^{-3} \pm 1.9x10^{-5}$ | |
| | **0.09** | $5.0x10^{-4} \pm 5.0x10^{-6}$ | $2.0x10^{-3} \pm 2.0x10^{-5}$ | $4.2x10^{-3} \pm 4.2x10^{-5}$ |
| | **0.1** | $5.0x10^{-4} \pm 5.0x10^{-6}$ | $2.0x10^{-3} \pm 2.0x10^{-5}$ | $4.3x10^{-3} \pm 4.3x10^{-5}$ |

**Table S4: Results of the numerical test of the semi-implicit 1D-uFDM stability condition.** Smallest values of $\boldsymbol{\beta}$ out of those tested ($\boldsymbol{\beta} = \mathbf{1x10^{-4}}, \mathbf{1x10^{-3}}, \mathbf{1x10^{-2}}$) for which simulations with $\Delta \boldsymbol{t}_{+}$ were unstable and $\Delta \boldsymbol{t}_{-}$ were stable. $\boldsymbol{D} = \mathbf{0.1}\ \boldsymbol{\mu m^2 s^{-1}}$.

| | | $\Delta x$ | | | | | | | | | |
|---|---|---|---|---|---|---|---|---|---|---|---|
| | | **0.01** | **0.02** | **0.03** | **0.04** | **0.05** | **0.06** | **0.07** | **0.08** | **0.09** | **0.1** |
| $\Delta y$ | **0.01** | 0.0001 | 0.0001 | 0.0001 | 0.0001 | 0.0001 | 0.0001 | 0.0001 | 0.0001 | 0.0001 | 0.0001 |
| | **0.02** | 0.0001 | 0.0001 | 0.0001 | 0.0001 | 0.0001 | 0.0001 | 0.0001 | 0.0001 | 0.0001 | 0.0001 |
| | **0.03** | 0.0001 | 0.0001 | 0.0001 | 0.0001 | 0.0001 | 0.0001 | 0.0001 | 0.0001 | 0.0001 | 0.0001 |
| | **0.04** | 0.0001 | 0.0001 | 0.0001 | 0.0001 | 0.0001 | 0.0001 | 0.0001 | 0.0001 | 0.0001 | 0.0001 |
| | **0.05** | 0.0001 | 0.0001 | 0.0001 | 0.0001 | 0.0001 | 0.0001 | 0.0001 | 0.0001 | 0.0001 | 0.0001 |
| | **0.06** | 0.0001 | 0.0001 | 0.0001 | 0.0001 | 0.0001 | 0.0001 | 0.0001 | 0.0001 | 0.0001 | 0.0001 |
| | **0.07** | 0.0001 | 0.0001 | 0.0001 | 0.0001 | 0.0001 | 0.0001 | 0.0001 | 0.0001 | 0.0001 | 0.0001 |
| | **0.08** | 0.0001 | 0.0001 | 0.0001 | 0.0001 | 0.0001 | 0.0001 | 0.0001 | 0.0001 | 0.0001 | 0.0001 |
| | **0.09** | 0.0001 | 0.0001 | 0.0001 | 0.0001 | 0.0001 | 0.0001 | 0.0001 | 0.0001 | 0.0001 | 0.0001 |
| | **0.1** | 0.0001 | 0.0001 | 0.0001 | 0.0001 | 0.0001 | 0.0001 | 0.0001 | 0.0001 | 0.0001 | 0.0001 |

**Appendix 4: 1D Fourier solution.**

Regarding notation: For the finite difference scheme, $u_n^\tau$ denotes the concentration of a species on mesh point $n$ at time $\tau$. To avoid confusion, here we define the notation for the Fourier solution. Let $v(x,t)$ or $v(x,y,t)$ describe the concentration of a species at position $x$ or $(x,y)$ at time $t$. The Fourier general solution will be comprised of the sum of fundamental solutions, using the principle of superposition. $v_n(x,t)$ or $v_n(x,y,t)$ will denote the $n^{\text{th}}$ fundamental solution. For this investigation, the 1D initial condition was $v(x,0) = e^{-x^2}$, 2D $v(x,y,0) = e^{-(x^2+y^2)}$ (Fig. S1a and b, $\alpha = 1$), and $D = 0.1\ \mu m^2 s^{-1}$.

The 1D Fourier solution informs the 2D solution so we will present that first. Consider a 1D space $-L$ to $L$, with periodic boundary conditions. Diffusive movement of $v(x,t)$ in the 1D space is given by the equation, $\frac{\partial}{\partial t} v(x,t) = D \frac{\partial^2}{\partial x^2} v(x,t)$, where $D$ is the diffusion coefficient. The periodic boundary conditions are,

$$v(-L,t) = v(L,t) \qquad \text{Boundary condition 1}$$

$$\frac{\partial}{\partial x} v(x,t)\Big|_{-L} = \frac{\partial}{\partial x} v(x,t)\Big|_{L} \qquad \text{Boundary condition 2}$$

The diffusion equation can be solved by separation of variables and Fourier series expansions (1, 2). Separation of variables states $v(x,t)$ will have the form, $v(x,t) = X(x)T(t)$. Substitute $v(x,t) = X(x)T(t)$ into the diffusion equation to get,

$$\frac{1}{D\,T(t)} \frac{\partial}{\partial t} T(t) = \frac{1}{X(x)} \frac{\partial^2}{\partial x^2} X(x) \qquad \text{Equation 3}$$

Both sides of the Equation 3 must equal a constant as one side depends only on time and the other depends only on space. Let the constant be $-\lambda$. The time aspect of Equation 3 is, $\frac{\partial}{\partial t} T(t) + \lambda\, D\, T(t) = 0$, which has the solution, $T(t) = Ce^{-\lambda D t}$, where $C$ is an arbitrary constant. The space aspect of Equation 3 is,

$$\frac{\partial^2}{\partial x^2} X(x) + \lambda\, X(x) = 0 \qquad \text{Equation 4}$$

There are three solutions to Equation 4, one for $\lambda < 0$, $\lambda = 0$ and $\lambda > 0$. The validity of these solutions are tested using boundary conditions 1 and 2. For $\lambda < 0$, $X(x) = k_1 \cosh\left(\sqrt{-\lambda}\, x\right) + k_2 \sinh\left(\sqrt{-\lambda}\, x\right)$, which has only trivial solutions. For $\lambda = 0$, $X(x) = k_1 x + k_2$. Applying Boundary Conditions 1 and 2 we get the fundamental solution,

$$v_0(x,t) = C_0 \qquad \text{Equation 5}$$

For $\lambda > 0$, Equation 4 has the solution, $X(x) = k_1 \cos\left(\sqrt{\lambda}\, x\right) + k_2 \sin\left(\sqrt{\lambda}\, x\right)$. Applying Boundary Condition 1 gives, $k_1 \cos\left(-\sqrt{\lambda}\, L\right) + k_2 \sin\left(-\sqrt{\lambda}\, L\right) = k_1 \cos\left(\sqrt{\lambda}\, L\right) + k_2 \sin\left(\sqrt{\lambda}\, L\right)$. As $\cos$ is even we get, $-k_2 \sin\left(\sqrt{\lambda}\, L\right) = k_2 \sin\left(\sqrt{\lambda}\, L\right)$. This equality holds if $k_2 = 0$, or $\lambda_n = \left(\frac{n\pi}{L}\right)^2$, $n = 1,2,3 \ldots$ . The solution $X(x) = k_1 \cos\left(\sqrt{\lambda}\, x\right) + k_2 \sin\left(\sqrt{\lambda}\, x\right)$ becomes $X_n(x) = k_{1n} \cos\left(\frac{n\pi}{L} x\right) + k_{2n} \sin\left(\frac{n\pi}{L} x\right)$, which satisfies Boundary Condition 2. Note the solution for $k_2 = 0$ is accounted for in $X_n(x)$ because of the principle of superposition. Thus we have the set of fundamental solutions,

$$v_n(x,t) = \left(A_n \cos\left(\frac{n\pi}{L} x\right) + B_n \sin\left(\frac{n\pi}{L} x\right)\right) e^{-\left(\frac{n\pi}{L}\right)^2 D t} \qquad \text{Equation 6}$$

where $A_n = k_{1n}C_n$ and $B_n = k_{2n}C_n$.
Using the principle of superposition, we write the general solution to the diffusion equation with Boundary Conditions 1 and 2 as a linear composition of all solutions (Equation 5 and Equation 6), $v(x,t) = v_0(x,t) + \sum_{n=1,2\ldots} v_n(x,t)$,

$$v(x,t) = C_0 + \sum_{n=1,2\ldots}\left[\left(A_n \cos\left(\frac{n\pi}{L}x\right) + B_n \sin\left(\frac{n\pi}{L}x\right)\right)e^{-\left(\frac{n\pi}{L}\right)^2 Dt}\right] \quad \text{Equation 7}$$

We use the initial condition and Fourier analysis to determine the coefficients of Equation 7, and thus obtain the solution of the diffusion equation with the given initial condition. The initial condition is $(x,0) = e^{-x^2}$. At $t = 0$ Equation 7 becomes a Fourier series, $e^{-x^2} = C_0 + \sum_{n=1,2\ldots}\left[A_n \cos\left(\frac{n\pi}{L}x\right) + B_n \sin\left(\frac{n\pi}{L}x\right)\right]$. The initial condition is an even function and so $B_n = 0 \; \forall \, n$. Fourier analysis states that $C_0 = \frac{1}{2L}\int_{-L}^{L} e^{-x^2}\,dx$, giving, and $A_n = \frac{1}{L}\int_{-L}^{L}\cos\left(\frac{n\pi}{L}x\right)e^{-x^2}\,dx$. Thus, the general solution for the 1D diffusion equation with initial condition $v(x,0) = e^{-x^2}$ and periodic boundary conditions is,

$$v(x,t) = C_0 + \sum_{n=1,2\ldots}\left[A_n \cos\left(\frac{n\pi}{L}x\right)e^{-\left(\frac{n\pi}{L}\right)^2 Dt}\right]$$
$$C_0 = \frac{1}{2L}\sqrt{\pi}\,\mathrm{erf}(L)$$
$$A_n = \frac{1}{2L}i\sqrt{\pi}\,e^{-\left(\frac{n\pi}{2L}\right)^2}\left(\mathrm{erfi}\left(\frac{n\pi}{2L} - iL\right) - \mathrm{erfi}\left(\frac{n\pi}{2L} + iL\right)\right)$$

**Appendix 5: 2D Fourier solution.**

Consider a 2D space $x \in [-L, L]$ and $y \in [-L, L]$ with periodic boundary conditions. The diffusion equation in 2D space is $\frac{\partial}{\partial t}v(x,y,t) = D\left(\frac{\partial^2}{\partial x^2}v(x,y,t) + \frac{\partial^2}{\partial y^2}v(x,y,t)\right)$, with periodic boundary conditions,

$$v(-L,y,t) = v(L,y,t)$$
$$v(x,-L,t) = v(x,L,t)$$
$$\left.\frac{\partial}{\partial x}v(x,y,t)\right|_{-L} = \left.\frac{\partial}{\partial x}v(x,y,t)\right|_{L}$$
$$\left.\frac{\partial}{\partial y}v(x,y,t)\right|_{-L} = \left.\frac{\partial}{\partial y}v(x,y,t)\right|_{L}$$

Separation of variables states $v(x,y,t)$ will have the form, $v(x,y,t) = X(x)Y(y)T(t)$. Substitute $v(x,y,t) = X(x)Y(y)T(t)$ into the 2D diffusion equation to get, $\frac{1}{D\,T(t)}\frac{\partial}{\partial t}T(t) = \frac{1}{X(x)}\frac{\partial^2}{\partial x^2}X(x) + \frac{1}{Y(y)}\frac{\partial^2}{\partial y^2}Y(y) = -\lambda = -(\lambda_n + \lambda_m)$. $T(t)$ and $X(x)$ have the same solution as they did when considering the 1D diffusion equation, and the solution for $Y(y)$ will have the same form as $X(x)$. However, for the 2D case $\lambda_n = \left(\frac{n\pi}{L}\right)^2$, $\lambda_m = \left(\frac{m\pi}{L}\right)^2$ and $\lambda = \left(\frac{n\pi}{L}\right)^2 + \left(\frac{m\pi}{L}\right)^2$.

$$T(t) = Ce^{-\left(\left(\frac{n\pi}{L}\right)^2 + \left(\frac{m\pi}{L}\right)^2\right)Dt}$$
$$X(x) = k_2$$
$$X_n(x) = k_{1n}\cos\left(\frac{n\pi}{L}x\right) + k_{2n}\sin\left(\frac{n\pi}{L}x\right)$$
$$Y(y) = k_4$$
$$Y_n(y) = k_{1m}\cos\left(\frac{m\pi}{L}y\right) + k_{2m}\sin\left(\frac{m\pi}{L}y\right)$$

The corresponding solutions for $v(x,y,t)$ are $v_{nm}(x,y,t)$,

$$v_{00}(x,y,t) = C_{00}$$
$$v_{0m}(x,y,t) = C_{0m}\left(A_m \cos\left(\frac{m\pi}{L}y\right) + B_m \sin\left(\frac{m\pi}{L}y\right)\right) e^{-\left(\frac{m\pi}{L}\right)^2 Dt}$$
$$v_{n0}(x,y,t) = C_{n0}\left(A_n \cos\left(\frac{n\pi}{L}x\right) + B_n \sin\left(\frac{n\pi}{L}x\right)\right) e^{-\left(\frac{n\pi}{L}\right)^2 Dt}$$
$$v_{nm}(x,y,t) = \left(A_n \cos\left(\frac{n\pi}{L}x\right) + B_n \sin\left(\frac{n\pi}{L}x\right)\right)\left(A_m \cos\left(\frac{m\pi}{L}y\right) + B_m \sin\left(\frac{m\pi}{L}y\right)\right) e^{-\left(\left(\frac{n\pi}{L}\right)^2+\left(\frac{m\pi}{L}\right)^2\right)Dt}$$

Using the principle of superposition, we write the general solution to the 2D diffusion equation with periodic boundary conditions,

$$v(x,y,t) = v_{00}(x,y,t) + \sum_{m=1,2\ldots} v_{0m}(x,y,t) + \sum_{n=1,2\ldots} v_{n0}(x,y,t) + \sum_{n,m=1,2\ldots} v_{nm}(x,y,t)$$

Again, we use Fourier analysis find the Fourier coefficients and solve the general solution for a given initial condition. The general solution for the 2D diffusion equation with initial condition $v(x,y,0) = e^{-(x^2+y^2)}$ is,

$$v(x,y,t) = C_{00} + \sum_{m=1,2\ldots} A_{0m} \cos\left(\frac{m\pi}{L}y\right) e^{-\left(\frac{m\pi}{L}\right)^2 Dt} + \sum_{n=1,2\ldots} A_{n0} \cos\left(\frac{n\pi}{L}x\right) e^{-\left(\frac{n\pi}{L}\right)^2 Dt} + \sum_{n,m=1,2\ldots} A_{nm} \cos\left(\frac{n\pi}{L}x\right) \cos\left(\frac{m\pi}{L}y\right) e^{-\left(\left(\frac{n\pi}{L}\right)^2+\left(\frac{m\pi}{L}\right)^2\right)Dt}$$
$$C_{00} = \frac{\pi}{(2L)^2}\mathrm{erf}(L)^2$$
$$A_{0m} = \frac{-\pi}{i(2L)^2}\mathrm{erf}(L)\, e^{-\left(\frac{m\pi}{2L}\right)^2}\left(\mathrm{erfi}\left(\frac{L}{i} - \frac{m\pi}{2L}\right) + \mathrm{erfi}\left(\frac{L}{i} + \frac{m\pi}{2L}\right)\right)$$
$$A_{n0} = \frac{-\pi}{i(2L)^2}\mathrm{erf}(L)\, e^{-\left(\frac{n\pi}{2L}\right)^2}\left(\mathrm{erfi}\left(\frac{L}{i} - \frac{n\pi}{2L}\right) + \mathrm{erfi}\left(\frac{L}{i} + \frac{n\pi}{2L}\right)\right)$$
$$A_{nm} = \frac{-\pi}{(2L)^2} e^{-\left(\frac{n\pi}{2L}\right)^2 - \left(\frac{m\pi}{2L}\right)^2}\left(\mathrm{erfi}\left(\frac{L}{i} - \frac{n\pi}{2L}\right) + \mathrm{erfi}\left(\frac{L}{i} + \frac{n\pi}{2L}\right)\right)\left(\mathrm{erfi}\left(\frac{L}{i} - \frac{m\pi}{2L}\right) + \mathrm{erfi}\left(\frac{L}{i} + \frac{m\pi}{2L}\right)\right)$$

**Appendix 6: Steady state accuracy of 1D-uFDMs.**

The accuracy of the 1D-uFDM solutions was investigated numerically by comparing it with the central slice of the 2D Fourier solution (Appendix 5). The comparisons were made using three sets of *in-silico* experiments. In set 1, we asked how 1D-uFDM accuracy was affected by mesh size by fixing $\Delta t$ and varying $\Delta x, \Delta y$. In set 2 we asked how accuracy was affected by increasing $\Delta t$. Here, we fixed $\Delta x, \Delta y$ and varied $\Delta t$. Finally, in set 3, we looked at the interdependence of $\Delta x, \Delta y$ and $\Delta t$ by varying all three.

First we looked at the accuracy of the 1D-uFDMs when predicting steady state on the central slice of the 2D Fourier solution. Simulation set 1: To ascertain how the accuracy of the 1D-uFDMs was affected by mesh size, for each $\Delta x = 0.01, 0.02, \ldots, 0.1\ \mu m$, a simulation was run for $\Delta y = 0.01, 0.02, \ldots, 0.1\ \mu m$. To ensure numerical stability for all $\Delta x, \Delta y$, and for both explicit and semi-implicit numerical methods, $\Delta t$ was set using the explicit 1D-uFDM stability condition (Fig. S3), $\Delta t_(\Delta x, \Delta y, D, \beta) = \Delta t_(0.01, 0.01, 0.1, 0.1) = \frac{1-\beta}{\mathrm{D}}\left(\frac{\Delta x^2 \Delta y^2}{2\Delta y^2 + \Delta x^2}\right) = 0.0003\ s$. The initial conditions were, $u_{1D}(x,t) = u_{1D}(x,0) = e^{-x^2}$, $u_{2D}(x,y,t) = u_{2D}(x,y,0) = e^{-(x^2+y^2)}$ (Fig. S1a and b, $\alpha = 1$). The protocol was thus, for each $\Delta x$, $\Delta y$ combination: The explicit and semi-implicit 1D-uFDMs were run to $300\ s$. The 2D Fourier solution was calculated at $0$ and $300\ s$. The mean squared distance (MSD) was calculated between each point on the 1D mesh and the corresponding point on the central row of the 2D mesh, for both $0$ and $300\ s$. If the MSD was zero at $0\ s$ then the MSD at $300\ s$ (steady state) was recorded. Changing mesh size did not dramatically affect the accuracy of the unrestricted finite difference methods, Fig. S4a. However, increasing $\Delta x$ resulted in a small increase in accuracy and increasing $\Delta y$ resulted in a small decrease. For $\Delta t_{-} = 0.0003\ s$ the semi-implicit unrestricted movement finite difference method was marginally more accurate than the explicit method.

Simulation set 2: Next, we investigated the effect of $\Delta t$ on the accuracy of the unrestricted movement finite difference methods by fixing $\Delta x = \Delta y = 0.1\ \mu m$ and varying $\Delta t$. $\Delta x = \Delta y = 0.1\ \mu m$ was chosen to give the largest numerically stable range of $\Delta t$ values . For both semi-implicit and explicit 1D-uFDMs $\Delta t = 0.0003$ to $0.03$ at intervals of 0.0027 s. As the semi-implicit 1D-uFDM is numerically stable for greater values of $\Delta t$ (Fig. S3b) the semi-implicit 1D-uFDM was also run for $\Delta t = 0.04, 0.05, \dots 0.1$ s (Fig. S4b). For the explicit 1D-uFDM increasing $\Delta t$ had no significant effect on the accuracy. The semi-implicit 1D-uFDM's accuracy was significantly increased with increasing $\Delta t$.

Given the difference in accuracy between the accuracy of semi-implicit and explicit 1D-FDMs for increasing $\Delta t$ one would expect that fixing $\Delta t >$ 0.0003 s and performing simulation set 1 again, would result in greater separation between semi-implicit and explicit 1D-uFDM accuracies than that shown in Fig. S4a. Simulations were run for $\Delta x \geq 0.05\ \mu m, \Delta y \geq 0.05\ \mu m$ and $\Delta t_{-}(0.5, 0.5, 0.1, 0.1) = \frac{1-\beta}{\mathrm{D}}\left(\frac{\Delta x^2 \Delta y^2}{2\Delta y^2 + \Delta x^2}\right) = 0.0075\ s$. The results of these simulations confirmed expectations (Fig. S4c). Fixing $\Delta x = \Delta y = 0.05\ \mu m$ and varying $\Delta t$ gave further confirmation of the accuracy trends reported in Fig. S4d.

For the explicit 1D-uFDM the choice of $\Delta x$, $\Delta y$ and $\Delta t$ has little effect on the accuracy of the numerical method, as long as the choice satisfices the explicit 1D-uFDM stability condition. For the semi-implicit 1D-uFDM increasing $\Delta t$ benefits accuracy, however to enable larger $\Delta t$ one must increase $\Delta y$ to ensure numerical stability. Simulation set 3 explores the interdependence of $\Delta x, \Delta y$ and $\Delta t$ with regards to steady state accuracy. For these simulations, all three parameters were changed. $\Delta x = 0.01, 0.02, \dots, 0.1\ \mu m$, $\Delta y = 0.01, 0.02, \dots, 0.1\ \mu m$, and $\Delta t$ was set for each $\Delta x$, $\Delta y$ combination using the equation $\Delta t_{-}(\Delta x, \Delta y, D, \beta) = \Delta t_{-}(\Delta x, \Delta y, 0.1, 0.1) = \frac{1-\beta}{\mathrm{D}}\left(\frac{\Delta x^2 \Delta y^2}{2\Delta y^2 + \Delta x^2}\right)\ s$ for explicit 1D-uFDM simulations and $\Delta t_{-}(\Delta y, D, \beta) = \Delta t_{-}(\Delta y, 0.1, 0.1) = \frac{1-\beta}{\mathrm{D}}\Delta y^2$ for semi-implicit. As expected, the explicit 1D-uFDM had the same accuracy with the larger values of $\Delta t$ as it did with $\Delta t = 0.0003\ \mathrm{s}$ (compare solid lines on Fig. S4a and S4e). As the stability boundary of the semi-implicit 1D-uFDM depends on $\Delta y$ only, increasing $\Delta y$ increases the value $\Delta t_{-}(\Delta y, 0.1, 0.1)$ and thus increases the accuracy of the semi-implicit numerical method (Fig. S4e). Continually increasing the value of $\Delta y$, and thus $\Delta t$, should eventually reduce the accuracy of the numerical method. To investigate this $\Delta x$ was set to $0.1\ \mu\mathrm{m}$, $\Delta y$ was increased from $0.01$ to $0.5\ \mu m$, and $\Delta t_{-}(\Delta y, 0.1, 0.1) = \frac{1-\beta}{\mathrm{D}}\Delta y^2\ s$. Indeed, the minimum MSD was found to be $3.2\mathrm{x}10^{-9}$ for $\Delta y = 0.38\ \mu m$, $\Delta t = 1.2996\ s$ (Fig. S4f).

Taken together these investigations show that to improve the accuracy of the steady state estimation when using the explicit 1D-uFDM one should choose a larger $\Delta x$ and smaller $\Delta y$, then set $\Delta t$ with the formula $\Delta t_{-}(\Delta x, \Delta y, D, \beta) = \frac{1-\beta}{\mathrm{D}}\left(\frac{\Delta x^2 \Delta y^2}{2\Delta y^2 + \Delta x^2}\right)$. For the semi-implicit method one should choose a larger $\Delta y$ and set $\Delta t$ using the formula $\Delta t_{-}(\Delta y, D, \beta) = \frac{1-\beta}{\mathrm{D}}\Delta y^2$.

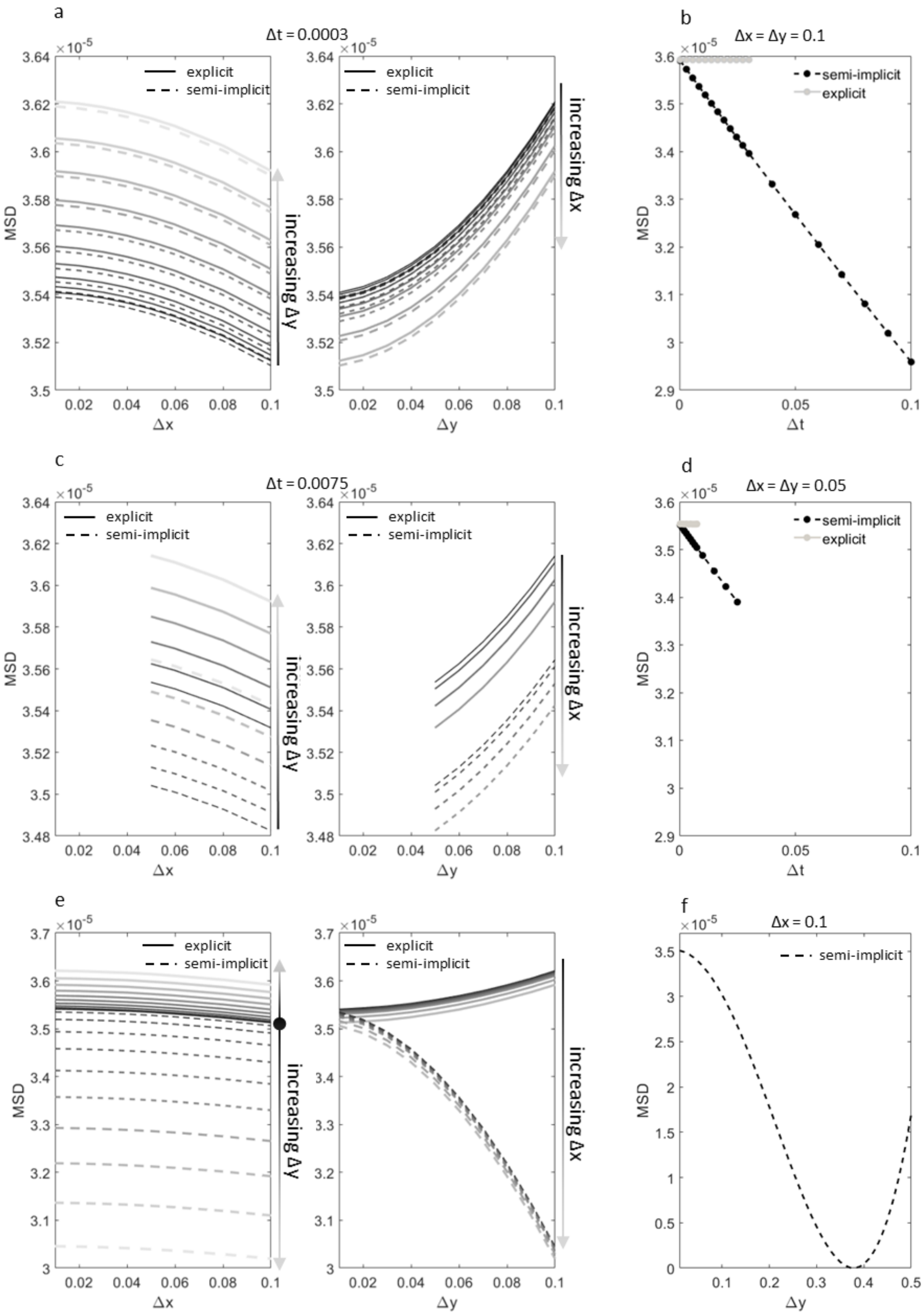

**Fig. S4: Accuracy of unrestricted movement finite difference methods at steady state.** Solid lines show the results of the explicit method, dotted the semi-implicit. In (a), (c) and (e) Lighter grey indicates increasing $\Delta x$ or $\Delta y$. **(a)** MSD between the explicit and semi-implicit 1D-FDMs and the 2D Fourier solutions at steady state, for $\boldsymbol{\Delta x = 0.01, 0.02, \dots, 0.1\ \mu m}$, $\boldsymbol{\Delta y = 0.01, 0.02, \dots, 0.1\ \mu m}$ and $\boldsymbol{\Delta t = 0.0003}$. **(b)** MSD at steady state for $\boldsymbol{\Delta x = \Delta y = 0.1\ \mu m}$, varying $\boldsymbol{\Delta t}$. **(c)** MSD between the explicit and semi-implicit 1D-FDMs and the 2D Fourier solutions at steady state, for $\boldsymbol{\Delta x \geq 0.05, \Delta y \geq 0.05}$ and $\boldsymbol{\Delta t = 0.0075}$. **(d)** MSD at steady state for $\boldsymbol{\Delta x = \Delta y = 0.5\ \mu m}$, varying $\boldsymbol{\Delta t}$. **(e)** MSD at steady state for $\boldsymbol{\Delta x = 0.01, 0.02, \dots, 0.1\ \mu m}$, $\boldsymbol{\Delta y = 0.01, 0.02, \dots, 0.1\ \mu m}$, and $\boldsymbol{\Delta t_{-} = \frac{1-\beta}{D}\left(\frac{\Delta x^2 \Delta y^2}{2\Delta y^2 + \Delta x^2}\right) s}$ for explicit simulations and $\boldsymbol{\Delta t_{-} = \frac{1-\beta}{D}\Delta y^2\ s}$ for semi-implicit. **(f)** MSD at steady state for semi-implicit 1D-uFDM, $\boldsymbol{\Delta x = 0.1\ \mu m}$, $\boldsymbol{\Delta y = 0.01, 0.02, \dots 0.5\ \mu m}$, and $\boldsymbol{\Delta t_{-}(\Delta y, 0.1, 0.1) = \frac{1-\beta}{D}\Delta y^2\ s}$.

**Appendix 7: Accuracy dynamics of 1D-uFDMs.**

Next, we compare the dynamics of the 1D-uFDMs to see how faithfully they estimate the approach to steady state. Similar to the steady state accuracy investigation we performed three sets of simulations. Simulation set 1: Investigate the time course dependence of 1D-uFDMs accuracy, and its relationship to mesh size. As calculating the Fourier solution every second for $300\,s$ is computationally time consuming we chose to investigate fewer mesh sizes than in the steady state accuracy investigation. For each $\Delta x = 0.02, 0.04, \dots, 0.1\ \mu m$, a simulation was run for $\Delta y = 0.02, 0.04, \dots, 0.1\ \mu m$. Again, to ensure numerical stability for all $\Delta x, \Delta y$, and for both explicit and semi-implicit numerical methods, $\Delta t$ was set using the explicit 1D-uFDM stability condition (Fig. S3). $\Delta t_{-}(\Delta x, \Delta y, D, \beta) = \Delta t_{-}(0.01, 0.01, 0.1, 0.1) = 0.0003\,s$. The initial conditions were $u_{1D}(x,t) = u_{1D}(x,0) = e^{-x^2}$, $u(x,y,t) = u(x,y,0) = e^{-(x^2+y^2)}$ (Fig. S1a and b, $\alpha = 1$). The protocol was thus, for each $\Delta x$, $\Delta y$ combination: The explicit and semi-implicit 1D-uFDMs were run to $300\,s$. As $\Delta t$ did not divide exactly into one, the 1D-uFDMs concentration profiles were recorded at close to one second intervals, and the times of data collection were recorded. The 2D Fourier solutions were calculated at the same times as the 1D-uFDMs' data was collected. For every time point the MSD was calculated between the 1D-uFDMs and the central row of the 2D Fourier solution.

Fig. S5a shows the MSD time evolution. As with the steady state MSD analysis, MSD dynamics were similar for all $\Delta x$, $\Delta y$ tested, for both the explicit and semi-implicit 1D-uFDMs. In all cases the MSD increased rapidly at around $30\,s$. To understand the cause of the rapid MSD increase we calculated the absolute distance between each point of the semi-implicit 1D-uFDM and the central row of the 2D Fourier solution, at every second, for $\Delta x = 0.1\,\mu m$, $\Delta y = 0.02\,\mu m$ as this had the smallest MSD at steady state (Fig. S4a, Fig. S5c). The absolute distance analysis showed that the absolute distance began to increase at the boundary edges and propagated through the mesh. Looking at the semi-implicit 1D-uFDM concentration profile we see that the increase in absolute distance at the boundaries corresponds to the rate of change of concentration at the boundaries (compare Fig. S5c and S5d, see also S5e). This is unsurprising as estimating concentrations $u^{\tau}_{\sqrt{(N/2\,\Delta x)^2+\Delta y^2}}$, which would lie at the boundaries of the 2D domain (black circle Fig. S5b), involves a 1D interpolation into space without any concentration information (orange circle Fig. S5b). As the curvature at the boundary increases so too does the interpolation error.

Looking more closely at the MSD time course data (Fig. S5a), we see a small peak before the edge error is detected (Fig. S6a and S6b). This initial MSD peak has a strong $\Delta y$ dependency. To ask if this initial MSD peak was a result of interpolation error we looked at the interpolation accuracy data (Fig. S2, row $\alpha = 1$, $\Delta x = 0.1\,\mu m$). This data showed that the interpolation accuracy does decrease with increasing $\Delta y$, and that the decrease is predominantly generated in the central region, around the location of the peak. However the size of the initial MSD peak is not accounted for by the interpolation inaccuracy on the initial concentration (Fig. S5a, row 2, column 1, $\alpha = 1$, $\Delta x = 0.1\,\mu m$, $L = 12\,\mu m$, divide max SSD in figure by $L/\Delta x = 120$ to get a maximum MSD around $10^{-13}$).

The 1D-uFDMs are composed of two parts, the 1D-FDM (highlighted in blue text in the equations below), and the terms utilising the interpolated concentrations.

$$u_n^{\tau+1} = u_n^\tau + d_x(u_{n-1}^\tau - 2u_n^\tau + u_{n+1}^\tau) + 2d_y\left(u^\tau_{\sqrt{(n\Delta x)^2+\Delta y^2}} - u_n^\tau\right)$$
$$u_n^{\tau+1} = u_n^\tau + d_x(u_{n-1}^{\tau+1} - 2u_n^{\tau+1} + u_{n+1}^{\tau+1}) + 2d_y\left(u^\tau_{\sqrt{(n\Delta x)^2+\Delta y^2}} - u_n^\tau\right)$$

To ask if the error inherent in the 1D-FDMs could account for the initial MSD peak the 1D-uFDMs, we compared the solutions of the 1D-FDMs with the 1D Fourier solutions for the diffusion equation (1D Fourier solution derived in Appendix 4). To enable direct error comparisons between the 1D-FDMs inherent error analysis and the 1D-uFDMs error analysis the 1D-FDMs were run for $300\ s$, with $\Delta x = 0.02, 0.04, \dots, 0.1\ \mu m$, $\Delta t = 0.0003\ s$ and initial condition $u_{1D}(x,t) = u_{1D}(x,0) = e^{-x^2}$. Again, as $\Delta t$ does not divide exactly into one, the 1D-FDMs concentration profiles were recorded at close to one second intervals, and the times of data collection were recorded. The 1D Fourier solutions were calculated at the same times. For every time point, the MSDs and SDs were calculated between the 1D-FDMs and the 1D Fourier solutions. Fig. S6c shows the 1D-FDMs MSD has initial MSD peaks comparable to the 1D-uFDMs initial MSD peaks for $\Delta y = 0.02, 0.04\ \mu m$, with the initial MSD peaks increasing in size with $\Delta x$ (compare Fig. S6a, S6b and S6c). As $\Delta y$ increases further interpolation inaccuracies are combined with the 1D-FDMs inherent error to increase the initial 1D-uFDMs MSD peaks, nonetheless these increased peak sizes remain within the same order of magnitude as the 1D-FDMs MSD peaks. To further confirm the hypothesis that the 1D-uFDMs initial MSD peaks could be attributed to the 1D-FDMs inherent error we compared 1D-FDMs and 1D-uFDMs SD kymographs of the first 20 s of SD data (Fig. S6d). Indeed, the 1D-uFDMs SD kymographs showed close resemblance to the 1D-uFDM SD kymographs for smaller $\Delta y$, with the largest SDs at the points of greatest curvature in the concentration profile. The last point of note is that the 1D-FDMs MSDs do not increase again after the initial peak (Fig. S6c) confirming that the dominant error increase in 1D-uFDMs MSD is due to interpolation inaccuracy at the domain edges (Fig. S5).

Simulation set 2: In the steady state analysis it was found that increasing $\Delta t$ improved the accuracy of the semi-implicit 1D-uFDM when estimating the homogeneous steady state, but not the explicit 1D-uFDM (Fig. S4b). To ask if the error inherent in the 1D-FDMs can account for the differences between the explicit and semi-implicit 1D-uFDM MSDs at steady state we performed a time course MSD analysis on 1D-FDMs and 1D-uFDMs with fixed $\Delta x = \Delta y = 0.1\ \mu m$ and increased $\Delta t$ from $0.0003\ s$ (the $\Delta t$ value in simulation set 1) to $0.03$ s for explicit 1D-FDM and 1D-uFDM and to $0.1$ s for the implicit 1D-FDM and semi-implicit 1D-uFDM, ensuring a range of values for $\Delta t$ while retaining numerical stability. We found that the error inherent in 1D-FDMs followed the same trend as did the steady state 1D-uFDMs MSD, namely increasing $\Delta t$ in the explicit 1D-FDM had no effect on the MSD, and increasing $\Delta t$ in the implicit 1D-FDM had a dramatic effect on the MSD (compare Fig. S4b and S6e). The 1D-FDM and 1D-uFDM initial MSD peaks followed the same trend, as expected (compare Fig. S6e and S6f inset). Looking at the entire MSD time course for the semi-implicit 1D-uFDM (Fig. S6f) we see that, for the values of $\Delta t$ chosen, the increase in MSD initial peak offsets the steady state MSD.

Simulation set 3: For completion we analysed the MSD dynamics in the semi-implicit 1D-uFDM for fixed $\Delta x = 0.1\ \mu m$, increasing $\Delta y = 0.01, 0.02, \dots 0.5\ \mu m$, and increasing $\Delta t_{-}(\Delta y, D, \beta) = \Delta t_{-}(\Delta y, 0.1, 0.1) = \frac{1-\beta}{\mathrm{D}} \Delta y^2\ s$ (steady state MSD shown in Fig. S4f). Increasing $\Delta y$ (and $\Delta t$) resulted in an increasing MSD initial peak, it did not have a minimum similar to the MSD at steady state, indicating the optimal $\Delta y$ for minimum MSD at steady state is found at a balance point between 1D-FDM error and interpolation boundary error (Fig. S6g).

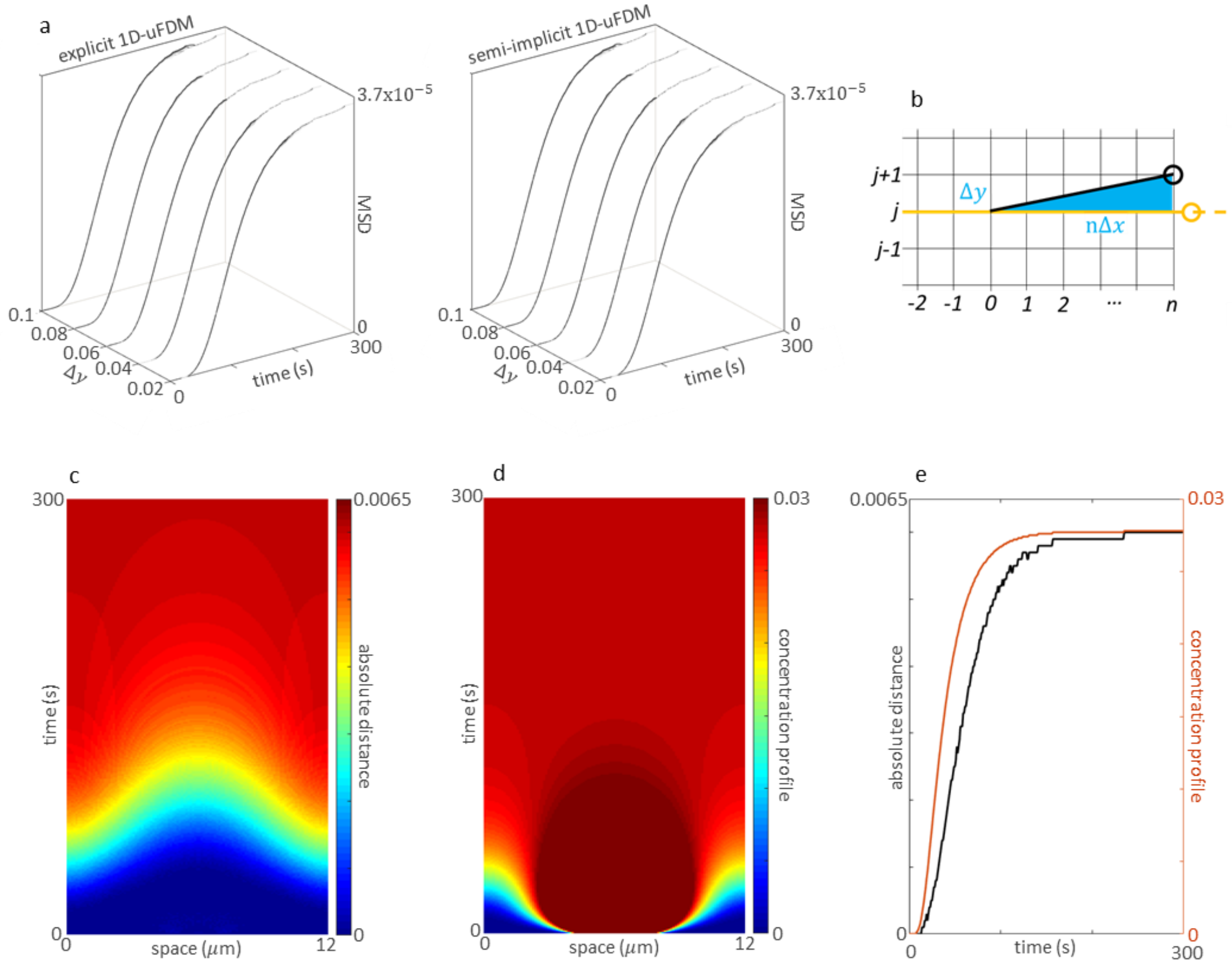

**Fig. S5: Accuracy dynamics of unrestricted movement finite difference methods: Interpolation error. (a)** MSD between the explicit and semi-implicit 1D-FDMs and the 2D Fourier solutions for $\Delta x = \mathbf{0.02, 0.04, \dots, 0.1}\ \boldsymbol{\mu m}$, $\Delta y = \mathbf{0.02, 0.04, \dots, 0.1}\ \boldsymbol{\mu m}$ and $\boldsymbol{\Delta t} = \mathbf{0.0003}\ \boldsymbol{s}$. For each $\boldsymbol{\Delta y}$ the five MSD curves for $\Delta x = \mathbf{0.02, 0.04, \dots, 0.1}\ \boldsymbol{\mu m}$ are drawn on top of each other. **(b)** Cartoon showing the estimation of $\boldsymbol{u}^{\tau}_{\sqrt{(N/2\ \Delta x)^2 + \Delta y^2}}$ at the boundary. **(c)** Kymograph of the absolute distance between each point of the semi-implicit 1D-uFDM and the central row of the 2D Fourier solution, at every second, for $\boldsymbol{\Delta x} = \mathbf{0.1}\ \boldsymbol{\mu m}$, $\boldsymbol{\Delta y} = \mathbf{0.02}\ \boldsymbol{\mu m}$, $\boldsymbol{\Delta t} = \mathbf{0.0003}\ \boldsymbol{s}$. (d) Kymograph, with upper threshold defined by the steady state concentration, of the concentration profile of the semi-implicit 1D-uFDM for $\boldsymbol{\Delta x} = \mathbf{0.1}\ \boldsymbol{\mu m}$, $\boldsymbol{\Delta y} = \mathbf{0.02}\ \boldsymbol{\mu m}$, $\boldsymbol{\Delta t} = \mathbf{0.0003}\ \boldsymbol{s}$. **(e)** Graph of the absolute distance and concentration profile at the boundary.

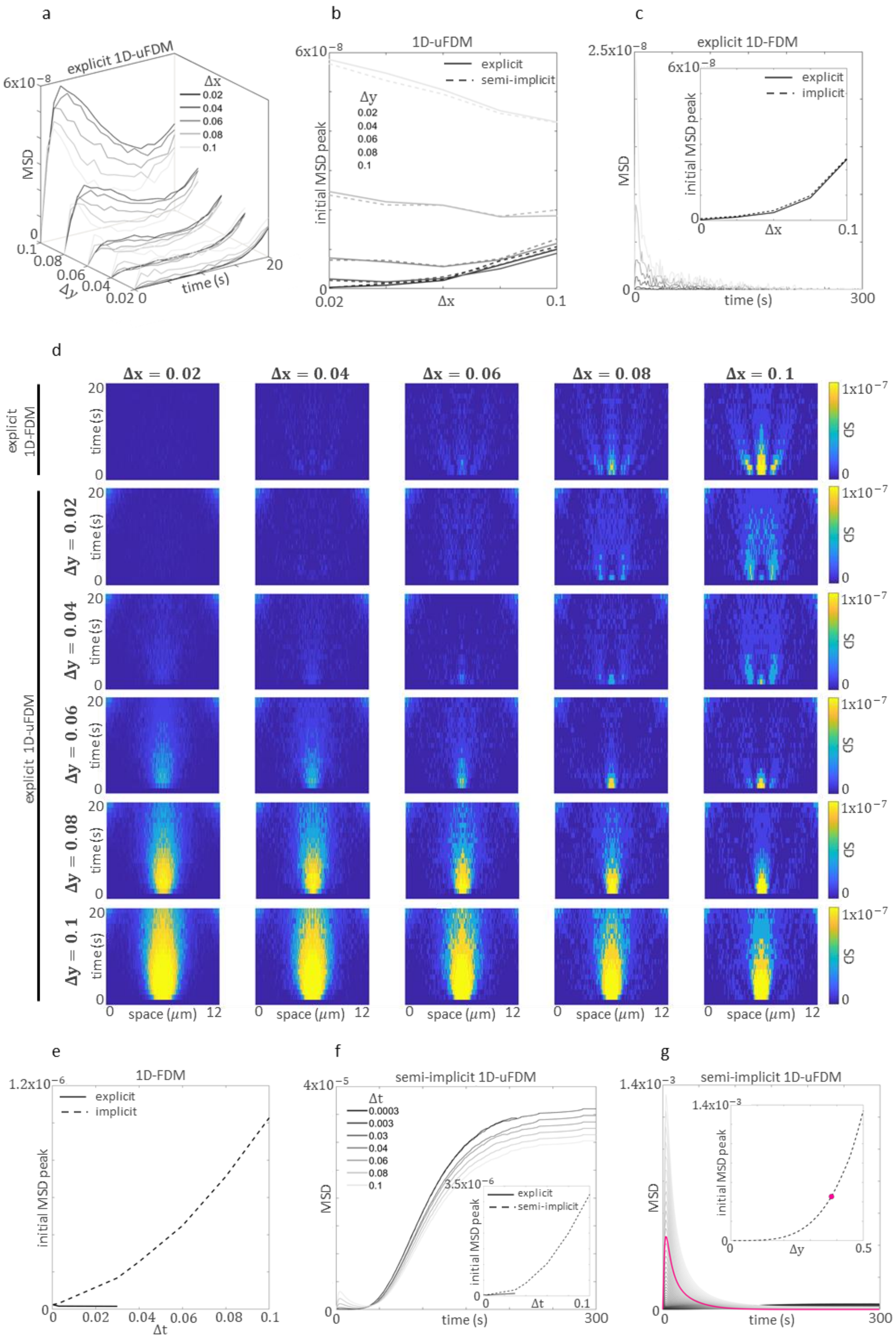

**Fig. S6: Accuracy dynamics of unrestricted movement finite difference methods: 1D-FDM error. (a)** Initial MSD peaks between the explicit 1D-uFDM and the 2D Fourier solutions, for $\Delta x = \mathbf{0.02, 0.04, \ldots, 0.1}\ \boldsymbol{\mu m}$, $\Delta y = \mathbf{0.02, 0.04, \ldots, 0.1}\ \boldsymbol{\mu m}$ and $\boldsymbol{\Delta t} = \mathbf{0.0003}\ \boldsymbol{s}$. **(b)** Values of initial MSD peaks for explicit and semi-implicit 1D-uFDM, for $\Delta x = \mathbf{0.02, 0.04, \ldots, 0.1}\ \boldsymbol{\mu m}$, $\Delta y = \mathbf{0.02, 0.04, \ldots, 0.1}\ \boldsymbol{\mu m}$ and $\boldsymbol{\Delta t} = \mathbf{0.0003}\ \boldsymbol{s}$. **(c)** MSD between the explicit 1D-FDM and the 1D Fourier solutions, for $\Delta x = \mathbf{0.02, 0.04, \ldots, 0.1}\ \boldsymbol{\mu m}$ and $\boldsymbol{\Delta t} = \mathbf{0.0003}\ \boldsymbol{s}$. Lighter grey indicates larger $\boldsymbol{\Delta x}$, key in Fig. S6a. **(c inset)** Values of initial MSD peaks for explicit and semi-implicit 1D-FDM. **(d)** SD kymographs showing the initial SD peak for $\Delta x = \mathbf{0.02, 0.04, \ldots, 0.1}\ \boldsymbol{\mu m}$, $\boldsymbol{\Delta t} = \mathbf{0.0003}\ \boldsymbol{s}$. Row 1 explicit 1D-FDM, rows 2 to 6 explicit 1D-uFDM for $\Delta y = \mathbf{0.02, 0.04, \ldots, 0.1}\ \boldsymbol{\mu m}$. **(e)** Initial MSD peaks for 1D-FDM, $\Delta x = \boldsymbol{\Delta y} = \mathbf{0.1}\ \boldsymbol{\mu m}$, varying $\Delta t$. **(f)** MSD for semi-implicit 1D-uFDM, $\Delta x = \boldsymbol{\Delta y} = \mathbf{0.1}\ \boldsymbol{\mu m}$, varying $\Delta t$. **(f inset)** Initial MSD peaks for 1D-uFDM, $\Delta x = \boldsymbol{\Delta y} = \mathbf{0.1}\ \boldsymbol{\mu m}$, varying $\Delta t$. **(g)** MSD for semi-implicit 1D-uFDM, $\Delta x = \mathbf{0.1}\ \boldsymbol{\mu m}$, $\Delta y = \mathbf{0.01, 0.02, \ldots 0.5}\ \boldsymbol{\mu m}$, and $\Delta \boldsymbol{t}_{-}(\boldsymbol{\Delta y}, \mathbf{0.1}, \mathbf{0.1}) = \frac{\mathbf{1-\beta}}{\mathbf{D}} \Delta \boldsymbol{y}^{\mathbf{2}}\ \boldsymbol{s}$. **(g inset)** Initial MSD peaks for semi-implicit 1D-uFDM, $\Delta x = \mathbf{0.1}\ \boldsymbol{\mu m}$, $\Delta y = \mathbf{0.01, 0.02, \ldots 0.5}\ \boldsymbol{\mu m}$, and $\Delta \boldsymbol{t}_{-}(\boldsymbol{\Delta y}, \mathbf{0.1}, \mathbf{0.1}) = \frac{\mathbf{1-\beta}}{\mathbf{D}} \Delta \boldsymbol{y}^{\mathbf{2}}\ \boldsymbol{s}$.

**Appendix 8: Accuracy of the semi-implicit 1D-uFDM when simulating diffusion.**

Fig. 2B-E shows the comparisons between the explicit 2D-FDM, 1D-FDM and 1D-uFDM when modelling diffusion in the full and reduced-dimension models. Fig. S7 shows the same investigation, but this time comparing the results of the explicit 2D-FDM, 1D-FDM and semi-implicit 1D-uFDM. $D = 0.1\ \mu m^2 s^{-1}$. A set of $\Delta x$, $\Delta y$ and $\Delta t$ values were chosen that gave an 'improved' steady state accuracy with a fairly small initial MSD peak, namely $\Delta x = \Delta y = 0.1\ \mu m$, $\Delta t = 0.1\ s$ (Fig. S6f). As expected setting $\Delta t = 0.01\ s$, the same as $\Delta t$ in the explicit comparison, gave a semi-implicit curve comparable to the explicit MSD curve (compare Fig. S7b with Fig. 2E).

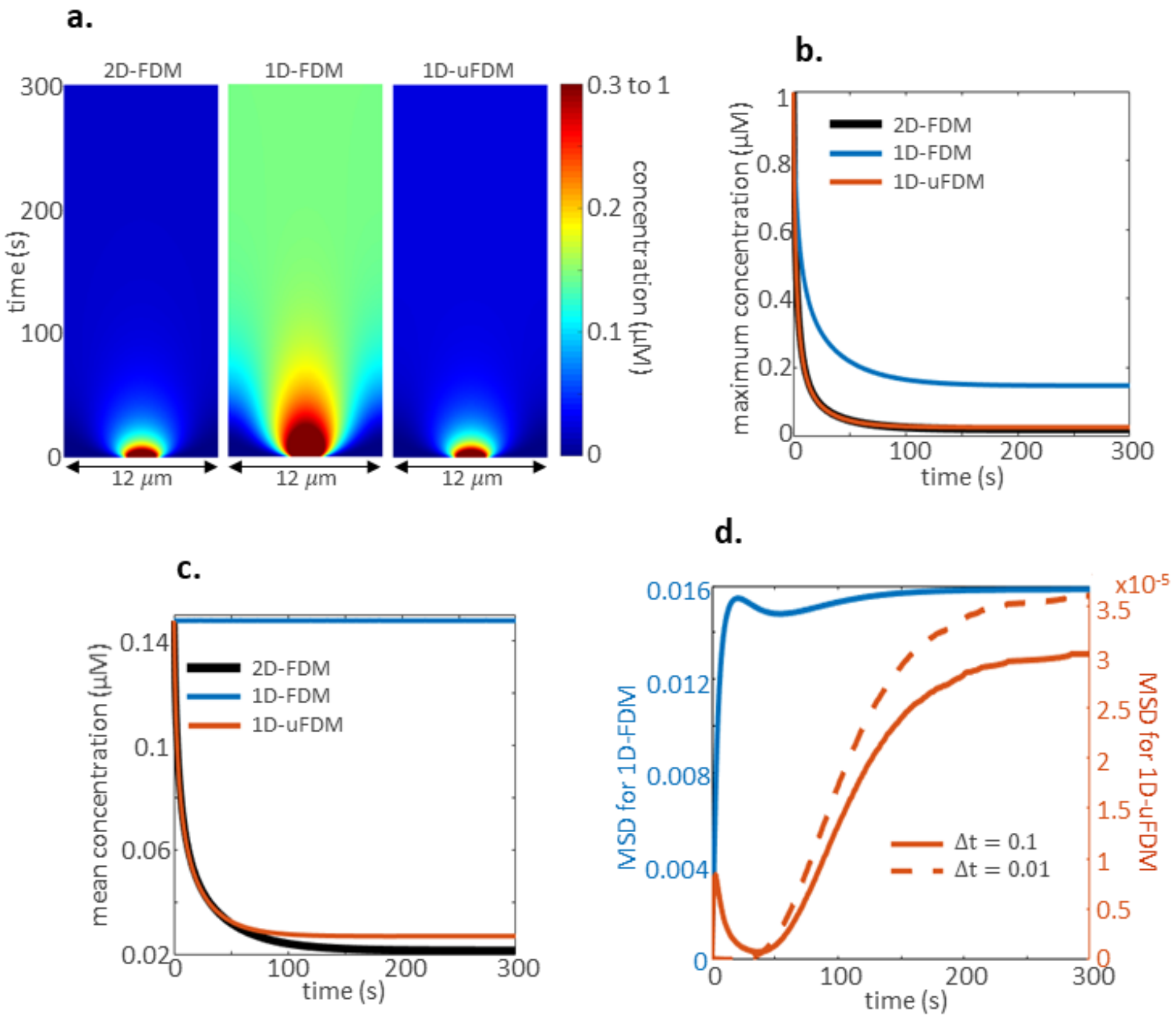

**Fig. S7: Accuracy dynamics of the implicit 1D-FDM and semi-implicit 1D-uFDMs when simulating diffusion in reduced dimension models. (a)** Kymographs of molecule $\boldsymbol{u}$ on the focal plane of the 2D model and in the 1D models. **(b)** Maximum concentration dynamics. **(c)** Mean concentration in the 1D solutions compared with the mean concentration in the focal plane of the 2D solution. **(d)** MSD between the focal plane in the 2D solution and the 1D solutions. 1D-FDM comparison shown in blue, 1D-uFDM comparison shown in red.

## Appendix 9: Accuracy of the 1D-FDM and 1D-uFDM when simulating FRAP.

The accuracy analysis for the 1D diffusion solutions when solving for an initial peak or an initial trough (i.e. the FRAP solution) is almost identical (compare Fig. 2C and E with Fig. S8b and c). Both the inherent 1D-FDM error and edge error (Appendix 7) increase proportionally with the concentration's curvature, and in both the peak and trough profiles the curvature is described by the same equation, $e^{-(x^2+y^2)}$ in the 2D case, $e^{-x^2}$ for 1D. Both peak and trough solutions are subject to the same diffusion coefficient so the curvature of the profiles change at the same rate giving the same accuracy analysis results.

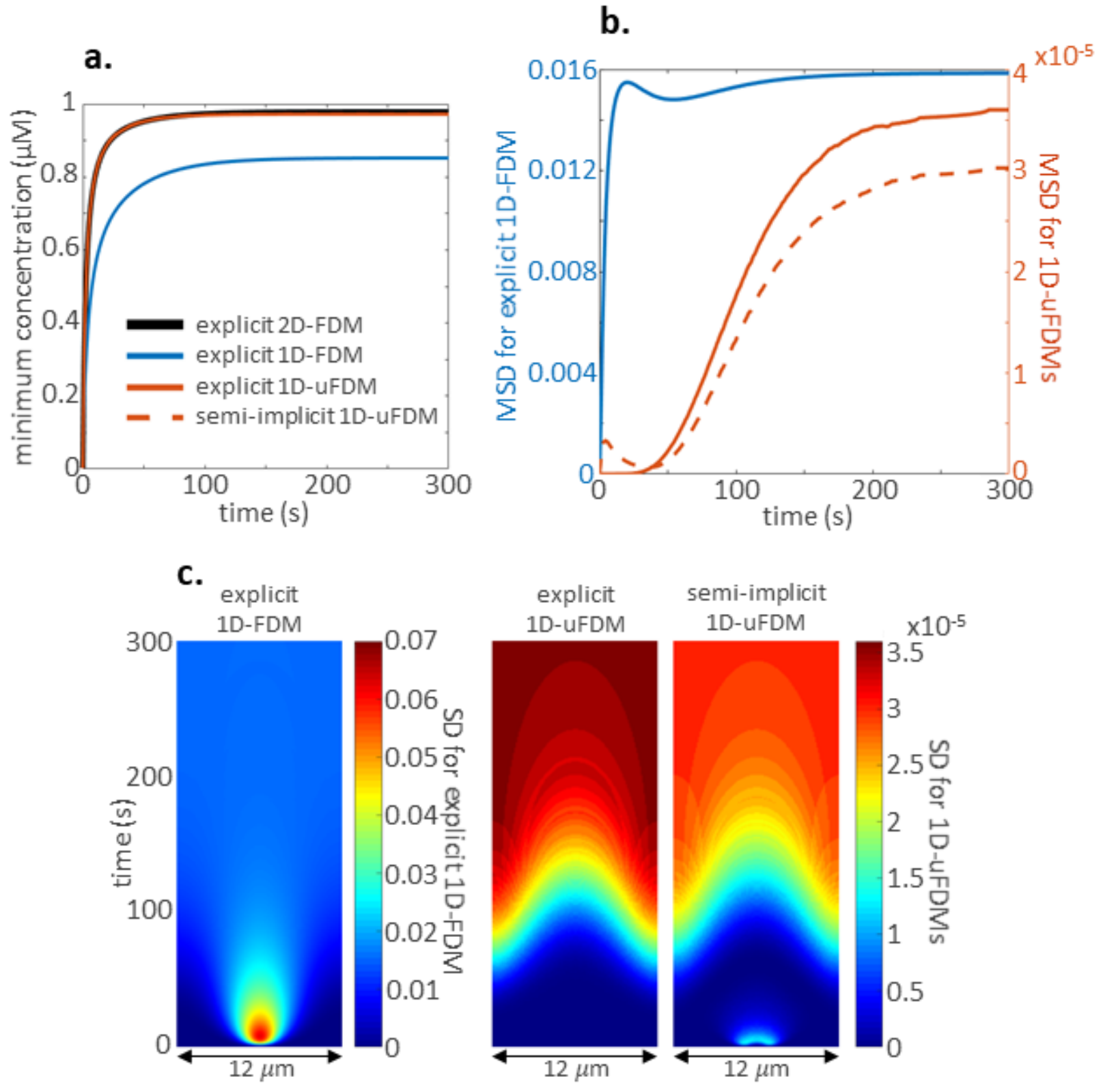

**Fig. S8: Accuracy dynamics of the 1D-FDM and 1D-uFDMs when simulating FRAP in reduced dimension models.** Figures support Fig. 3B-E. **(a)** Time course of the minimum concentration, i.e. the concentration at the centre of the ROI. **(b)** The MSD between the explicit 2D solution and the explicit 1D-FDM, explicit 1D-uFDM and semi-implicit 1D-uFDM. **(c)** Kymographs of the SD between the explicit 2D solution and the explicit 1D-FDM, explicit 1D-uFDM and semi-implicit 1D-uFDM.

**Appendix 10: Reaction-diffusion comparisons and the effect of geometry.**

The explicit 1D-uFDM captures RD dynamics through the focal plane more accurately than the explicit 1D-FDM (Fig. S9), with an MSD an order of magnitude smaller than that of the 1D-FDM for all $\alpha$ (Fig. S9c and f). Analysis of implicit 1D-FDM and semi-implicit 1D-uFDM comparisons gave almost identical results to the explicit comparisons, data not shown.

Recall, if the zero flux assumption holds, then the 1D-FDM solution is an accurate approximation of molecular dynamics through the focal plane of the 2D membrane. Thus the 1D-FDM solution to the RD equations is accurate if we were investigating an initial pulse on the surface of a spherical cell, for example (Fig. 1A, E). Fig. S10 shows the RD equations, $\alpha = 3$, solved on the body of an elongated cell using the square mesh and on a spherical cell using a spherical mesh. The same RD equations produce a patch on the square mesh and a ring on the spherical mesh.

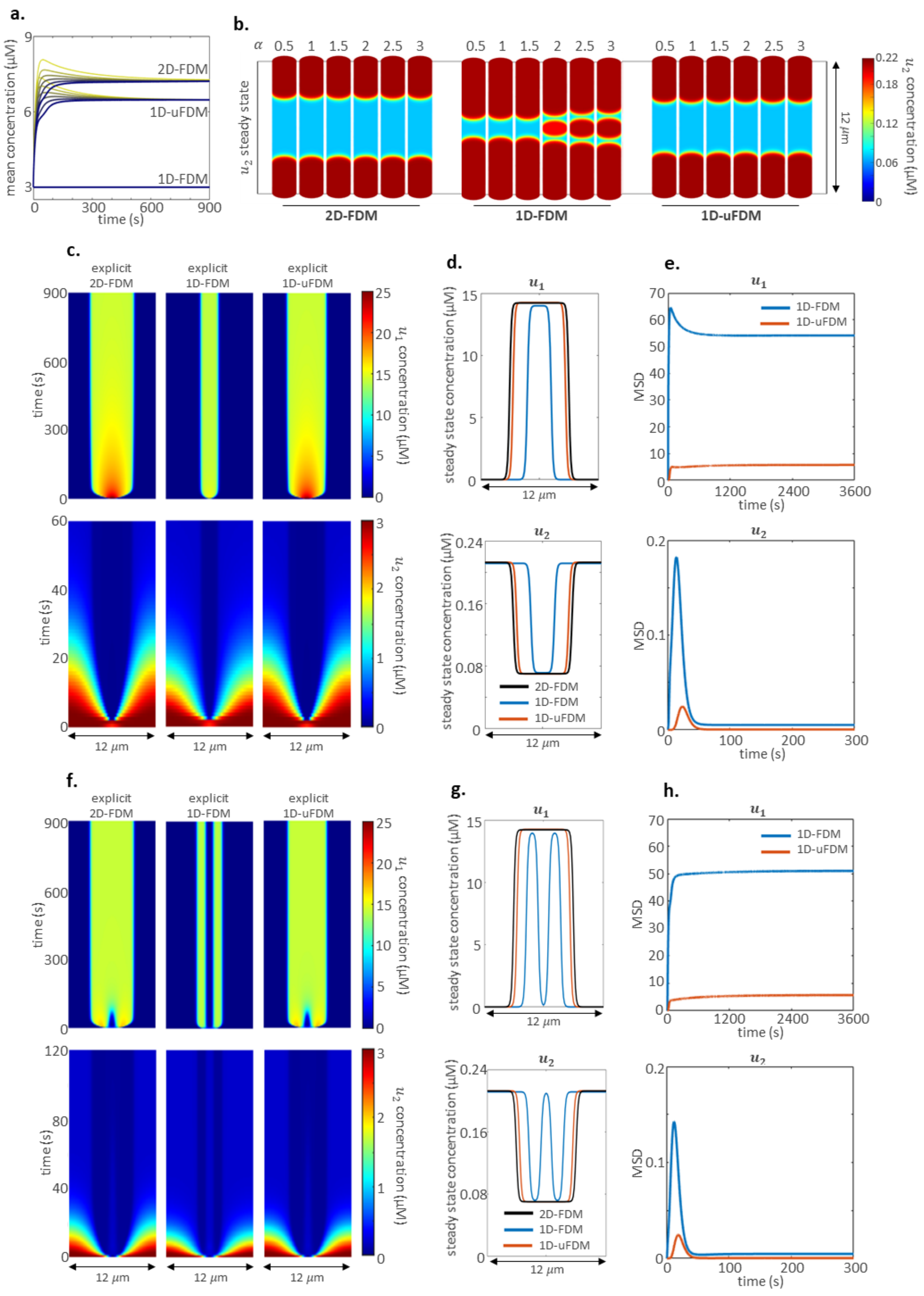

**Fig. S9: Accuracy dynamics of the 1D-FDM and 1D-uFDMs when solving RD equations.** Figures support Fig. 4B-D. **(a)** Mean concentration in the 1D solutions compared with the mean concentration in the focal plane of the 2D solution. **(b)** $\boldsymbol{u_2}$ steady state colorplots for the 2D solution on the focal plane, and 1D solutions. $\boldsymbol{\alpha = 0.5}$ in panels (c), (d) and (e). **(c)** Kymographs of the $\boldsymbol{u_1}$

and $u_2$ concentrations in the focal plane of the explicit 2D solution and 1D solutions. **(d)** $u_1$ and $u_2$ concentration profiles at steady state. **(e)** $u_1$ and $u_2$ MSD dynamics, comparing the concentrations in the focal plane of the 2D solution and the 1D solutions. $\alpha = 3$ in panels (f), (g) and (h). **(f)** Kymographs of the $u_1$ and $u_2$ concentrations in the focal plane of the 2D solution and 1D solutions. **(g)** $u_1$ and $u_2$ concentration profiles at steady state. **(h)** $u_1$ and $u_2$ MSD dynamics, comparing the concentrations in the focal plane of the 2D solution and the 1D solutions.

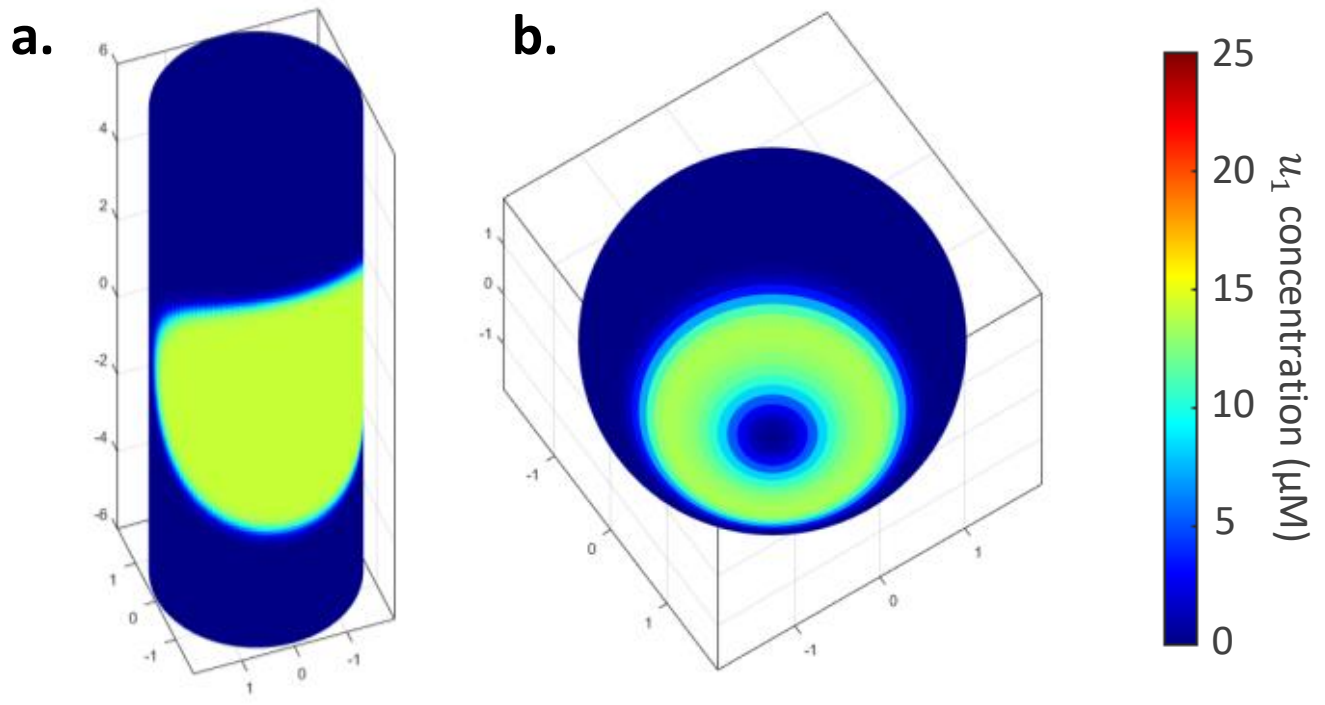

**Fig. S10: The effect of geometry on the RD solutions. Compare with Fig. 4C-D, $\alpha = 3$. (a)** Final steady state of the 2D-FDM solution to the RD equations, initial condition $\alpha = 3$, solved on the body of an elongated cell. **(b)** Final steady state of the 2D-FDM solution to the RD equations, initial condition $\alpha = 3$, solved on a spherical mesh using finite differences.

## SI References